\documentclass[useAMS,usenatbib]{mn2e}
\usepackage{graphicx}
\usepackage{latexsym}
\usepackage{psfig}
\usepackage{amssymb,amsmath}
\usepackage{subfig}
\usepackage{caption}

\title{Ionization--induced star formation IV: Triggering in bound clusters}

\author[J. E. Dale, B. Ercolano, I.A. Bonnell]{J. E. Dale$^{1}$\thanks{E-mail: dale@usm.lmu.de (JED)}, B. Ercolano$^{1}$, I. A. Bonnell$^{2}$\\
$^{1}$Excellence Cluster `Universe', Boltzmannstr. 2, 85748 Garching, Germany.\\
$^{2}$Department of Physics and Astronomy, University of St Andrews, North Haugh, St Andrews, Fife KY16 9SS}

\begin{document}
 
\pagerange{\pageref{firstpage}--\pageref{lastpage}} \pubyear{2006}

\maketitle

\label{firstpage}

\def\mnras{MNRAS}
\def\apj{ApJ}
\def\aj{AJ}
\def\aap{A\&A}
\def\apjl{ApJL}
\def\apjs{ApJS}
\def\araa{ARA\&A}
\def\pasj{PASJ}
 
\begin{abstract}
We present a detailed study of star formation occurring in bound star--forming clouds under the influence of internal ionizing feedback from massive stars across a spectrum of cloud properties. We infer which objects are triggered by comparing our feedback simulations with control simulations in which no feedback was present. We find feedback always results in a lower star--formation efficiency and usually but not always results in a larger number of stars or clusters. Cluster mass functions are not strongly affected by feedback, but stellar mass functions are biased towards lower masses. Ionization also affects the geometrical distribution of stars in ways that are robust against projection effects, but may make the stellar associations more or less subclustered depending on the background cloud environment. We observe a prominent pillar in one simulation which is the remains of an accretion flow feeding the central ionizing cluster of its host cloud and suggest that this may be a general formation mechanism for pillars such as those observed in M16. We find that the association of stars with structures in the gas such as shells or pillars is a good but by no means foolproof indication that those stars have been triggered and we conclude overall that it is very difficult to deduce which objects have been induced to form and which formed spontaneously simply from observing the system at a single time.
\end{abstract}

\begin{keywords}
stars: formation
\end{keywords}
 
\section{Introduction}
The influence of feedback from stars on the star--formation process itself is a long--standing and intriguing problem. Such feedback is usually invoked in the negative sense of `self--regulating star formation' -- the disruption of giant molecular clouds (GMCs) and embedded clusters by massive stars and the consequent shutting down of star formation. However, stellar feedback also has a positive component in the sense of triggered star formation -- the inducement of GMCs by massive stars to form new stars that they would not otherwise give birth to (\cite{1977ApJ...214..725E,1995ApJ...451..675E}, and see \cite{2011EAS....51...45E} for a brief up--to--date review). It is highly likely that stellar feedback operates in both modes simultaneously and that it triggers the formation of additional stars in some regions of GMCs while expelling gas from others. The interesting question is then whether its \textit{overall} effects are positive or negative.\\
\indent Observations of triggered star formation are legion and are usually loosely divided according to two popular models. The collect--and--collapse model involves the fragmentation, via gravitational and other instabilities, of a shell of dense material swept up by an expanding feedback--driven bubble. If simplifying assumptions about the bubble geometry and the smoothness of the background gas are made, this process lends itself easily to analytical \citep[e.g.][]{1994A&A...290..421W,2001A&A...374..746W} and numerical \citep[e.g.][]{2007MNRAS.375.1291D} study. There is also a large and growing body of observational work on this topic \citep[e.g.][]{2006A&A...446..171Z,2006A&A...458..191D,2008A&A...482..585D,2010A&A...518L..81Z} with which to compare the theoretical work. The case for triggered star formation in these systems is compelling. However, recent work by \cite{2011arXiv1109.3478W} questions whether the existence of a smooth shell is necessarily a pointer to the collect and collapse process in action. They perform simulations of HII regions expanding into fractal molecular clouds with various fractal dimensions and find that shell--like structures can readily be produced in even quite strongly non--uniform background clouds and reflect the initial gas distribution, not the collect and collapse process.\\
\indent If the assumptions of relatively simple (usually spherical) geometry and homogeneous ambient gas are dropped, as they must be in turbulent and highly non--uniform GMCs, identification of triggered stars becomes rather more difficult. Expanding HII regions (and wind/supernova bubbles) then encounter pre--existing structures in the surrounding gas, which may or may not be gravitationally unstable already and which will either be destroyed or induced to collapse. The latter outcome is described by the radiation--driven--implosion model (studied intensively by, e.g. \cite{1994A&A...289..559L,2003MNRAS.338..545K,2009MNRAS.393...21G,2011ApJ...736..142B}), although a stellar wind or supernova shock may have the same eventual result \citep[e.g.][]{1996ApJ...468..784F,1998ApJ...508..291V}. Disentangling triggered from spontaneous star formation (i.e. star formation that was going on anyway) in these circumstances becomes very difficult. The association of young stellar objects (YSOs) with shells/cavities \citep[e.g.][]{2003ApJ...595..900K,2008ApJ...688.1142K,2009A&A...503..107P} or pillars \citep[e.g.][]{1999AJ....117..225W,2005AJ....129..888S,2007ApJ...654..347L}, their proximity to ionization fronts \citep[e.g.][]{2009ApJ...700..506S} and the existence of bright--rimmed clouds \citep[e.g.][]{1995ApJ...455L..39S,2009A&A...497..789U} have all been used to infer star formation induced by stellar feedback but in all of these objects, the gas morphology and distribution of stars are very complex and difficult to interpret.\\
\indent Triggering may also be inferred more generally by searching for instances of sequential or self--propagating star formation, which can in principle be inferred by looking for spatial age gradients in star--forming regions or complexes. This idea was first proposed by \cite{1978SvAL....4...66E} and was confirmed on large scales (hundreds of pc to $\sim$1 kpc) by \cite{1989SvAL...15..388S}. On the scale of single associations, \cite{1995ApJ...451..675E} did the seminal theoretical work and considerable observational work followed. \citep[e.g.][]{1985ApJ...297..599D,2005AJ....129..776N,2006PASJ...58L..29M,2010ApJ...713..883B}. Recently, attempts have been made to look for statistical correlations between the positions of young stellar objects and infra--red bubbles. \cite{2012MNRAS.421..408T} and \cite{2012arXiv1203.5486K} both find statistically--significant overdensities of YSOs within and especially on the borders of, young feedback--driven bubbles. These groups infer that a few tens of percent of all massive stars in the Milky Way may have been triggered.\\
\indent In two previous papers, \cite{2007MNRAS.375.1291D}, \cite{2012MNRAS.tmp.2723D}, we have attempted to contribute to this discussion by modelling the effect on star formation in (unbound and bound, respectively) turbulent GMCs of external radiation by an arbitrarily--placed O--star. By comparing with control runs in which feedback was absent, we were able to quantify the degree of triggering, which we found to be modest (increasing the star--formation efficiency by $\sim30\%$ at most) in both cases. In this paper, we seek to model triggering in a more realistic setting and study the influence on the star formation in GMCs by stars that have already formed within that cloud. We take as our starting point a series of simulations described in \cite{2012MNRAS.424..377D}, hereafter Paper 1, whose purpose was to assess how efficient O--star photoionization alone could be in disrupting bound GMCs.\\
\indent In Paper 1, we describe in some detail the overall dynamical reaction of our model clusters to the ionizing feedback of the O--stars or O--star--hosting clusters formed within. Models C, G and H form no stars at all. Models E, F, B and X form stars/clusters vigorously but a combination of dense gas and strong accretion flows stifling the ionization of fresh gas, and the large escape velocities of these systems reduced the impact of ionization severely. In contrast, runs A, D, I and J were strongly affected by feedback, with several tens of percent of their gas reserves being expelled in the canonical 3Myr time window before the first supernova explosions. However, we noted that in the none of the clouds was feedback able to bring star formation to a halt, although star formation was noticeably slowed in the runs where feedback had tangible effects. However, these simulations also exhibited many morphological features that are often taken to be signposts of triggered star formation. We suggested in Paper 1 that the negative impact of gas expulsion might be to some extent counterbalanced by triggering of star formation -- i.e. by the birth of stars/clusters in the feedback simulations which would not have formed in the absence of feedback. In this paper, we investigate this possibility in detail by performing control runs identical to runs A, D, I and J in all respects except that photoionization was forbidden. We find that triggered star/cluster formation occurs in all our simulations, but that the overall effect of feedback on the star--formation efficiency is always negative..
\section{Numerical methods}
\begin{figure}
\includegraphics[width=0.5\textwidth]{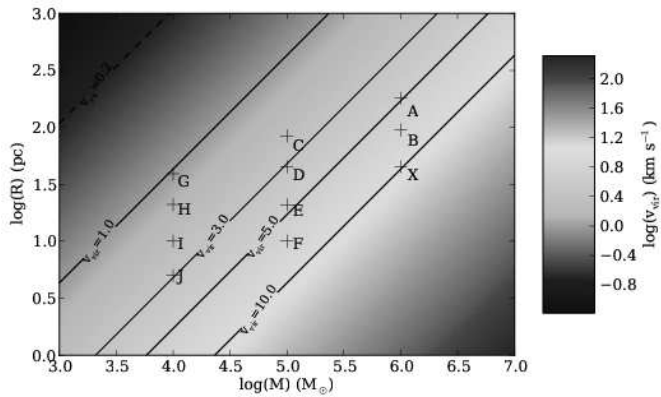}
\caption{Cluster mass--radius parameter space studied in Paper 1 and this work. Colours and black contour lines are velocities required to give uniform clusters of given mass and radius our chosen virial ratio of 0.7.} 
\label{fig:paramspace}
\end{figure}

Our numerical methods are identical to those detailed in Paper 1 and we will describe them only very briefly here. We use a hybrid N--body/SPH code based on that described by Benz \citep{1990nmns.work..269B}, updated to model star formation using the sink particle technique \citep{1995MNRAS.277..362B} and with an algorithm to simulate photoionization from multiple point sources \citep{2007MNRAS.382.1759D}, Dale and Ercolano (2012), in prep.\\
\indent The cold neutral gas is treated using a piecewise barotropic equation of state from \cite{2005MNRAS.359..211L}, defined so that $P = k \rho^{\gamma}$, where
\begin{eqnarray}
\begin{array}{rlrl}
\gamma  &=  0.75  ; & \hfill &\rho \le \rho_1 \\
\gamma  &=  1.0  ; & \rho_1 \le & \rho  \le \rho_2 \\
\gamma  &=  1.4  ; & \hfill \rho_2 \le &\rho \le \rho_3 \\
\gamma  &=  1.0  ; & \hfill &\rho \ge \rho_3, \\
\end{array}
\label{eqn:eos}
\end{eqnarray}
and $\rho_1= 5.5 \times 10^{-19} {\rm g\ cm}^{-3} , \rho_2=5.5 \times10^{-15} {\rm g cm}^{-3} , \rho_3=2 \times 10^{-13} {\rm g\ cm}^{-3}$. The thermodynamics are taken to be dominated by line cooling at low densities, dust cooling at intermediate densities optically--thick heating at high densities, with a final isothermal phase to permit sink particle formation. This choice of equation of state is extensively discussed justified and tested in Paper 1. Our initially--smooth model clouds are seeded with a Kolmogorov turbulent velocity such that the clouds are bound. For convenience, we reproduce here Figure \ref{fig:paramspace} from Paper 1, depicting the mass, radius and RMS turbulent velocities of the clouds we chose to model, and also in Table \ref{tab:params} the relevant lines from Table 1 in Paper 1 giving the basic parameters of Runs A, D, I and J.\\
\begin{table*}
\begin{tabular}{|l|l|l|l|l|l||l|l|l|}
Run&Mass (M$_{\odot}$)&Radius (pc)&v$_{\rm RMS}$ (km s$^{-1}$)&$\langle$ n(H$_{2}$) $\rangle$ (cm$^{-3}$)&t$_{\rm ff}$ (Myr)&$\langle$ T(K) $\rangle$&(E$_{\rm kin}$+E$_{\rm therm}$)/$|$E$_{\rm pot}|$\\
\hline
A&$10^{6}$&180&5.0&2.9&19.6&143&0.72\\
\hline
D&$10^{5}$&45&3.0&15&7.70&92&0.78\\
\hline
I&$10^{4}$&10&2.1&136&2.56&53&0.79\\
\hline
J&$10^{4}$&5&3.0&1135&0.90&32&0.72\\
\end{tabular}
\caption{Initial properties (mass, radius, turbulent velocity dispersion, mean initial molecular number density, freefall time, mean initial temperature and virial ratio) of Runs A, D, I and J.}
\label{tab:params}
\end{table*}
\indent Sink particles represent stars or clusters depending on mass resolution and are given ionizing luminosities if they are sufficiently massive as detailed in Paper 1. In the simulations where sink particles represent clusters, sinks approaching each other to within their accretion radii are merged if they are bound to one another.\\
\indent As in Paper 1, clouds are evolved without feedback until three stars or clusters sufficiently massive to possess ionizing fluxes are present. Simulations were then run for as close to 3Myr after the initiation of ionization as was numerically practicable, although in the case of Run J, which forms a rather dense cluster, we were obliged to halt the simulation after only $\approx$1.3 Myr. Despite this, the evolution of the ionized and control runs is plainly very different over this timescale.\\
\section{Results}
\subsection{Changes in gas structures due to feedback}
\begin{figure*}
     \centering
     \subfloat[Run A]{\includegraphics[width=0.45\textwidth]{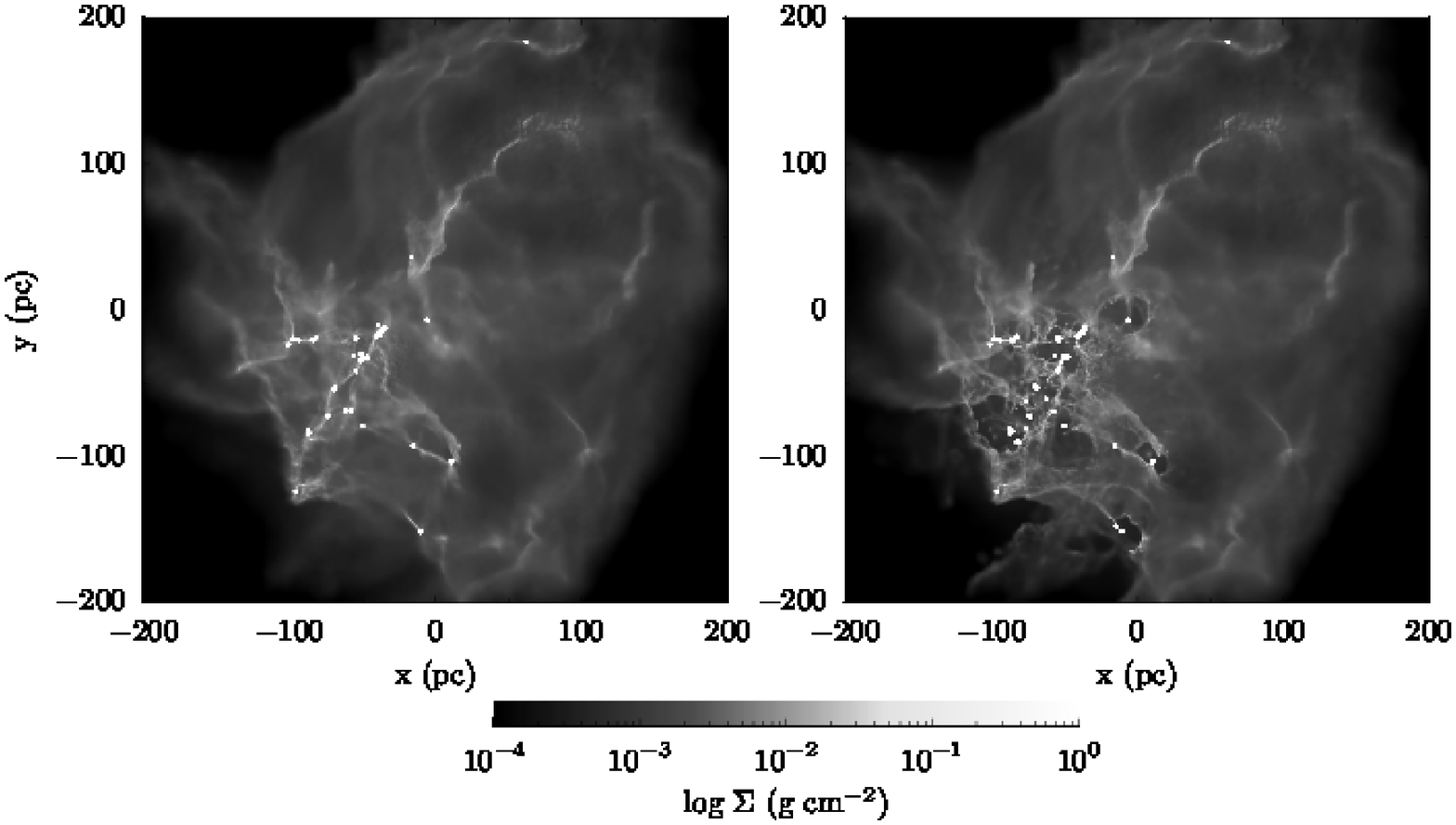}}     
     \hspace{.1in}
     \subfloat[Run D]{\includegraphics[width=0.45\textwidth]{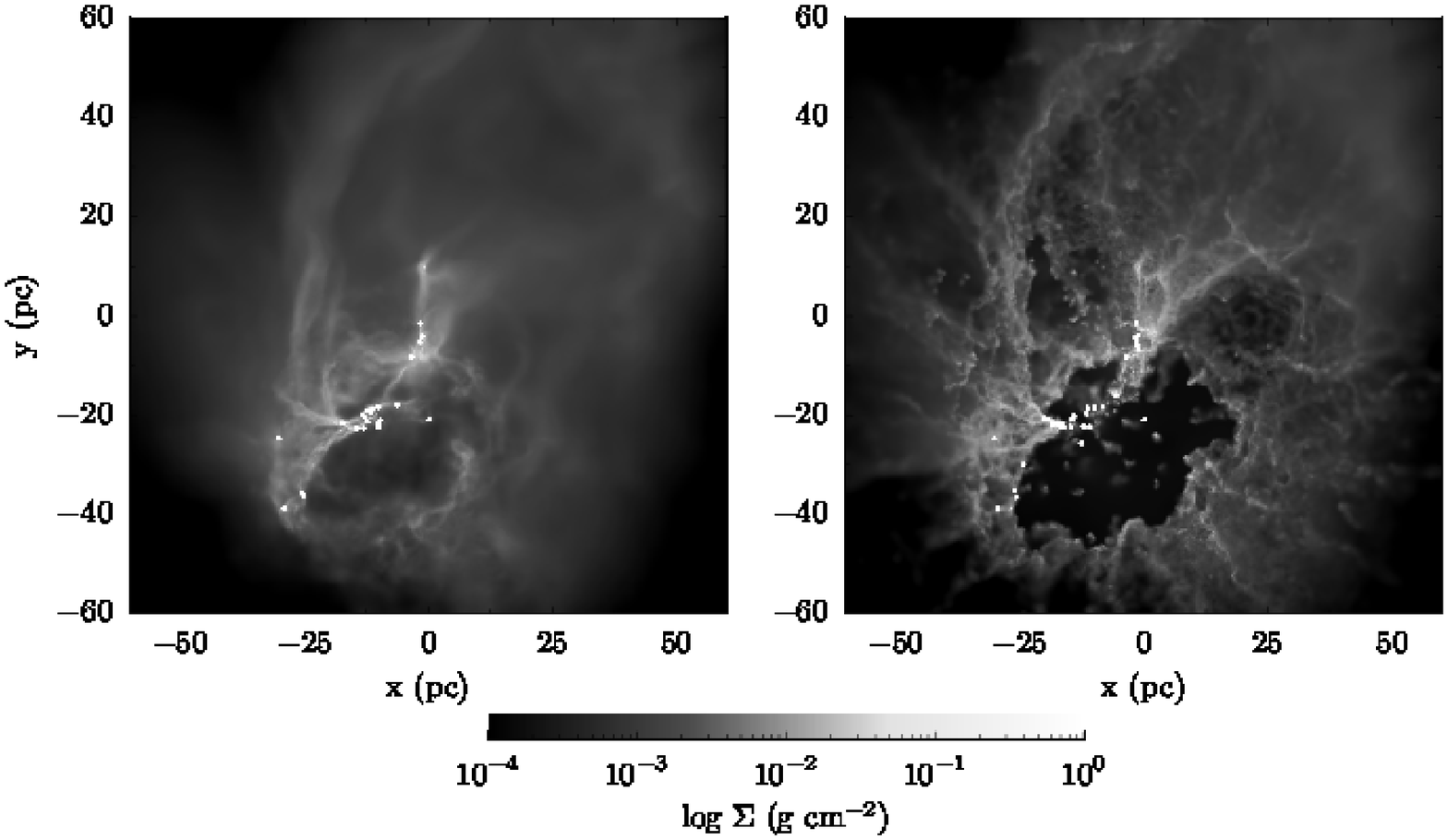}}
     \vspace{.1in}
     \subfloat[Run I]{\includegraphics[width=0.45\textwidth]{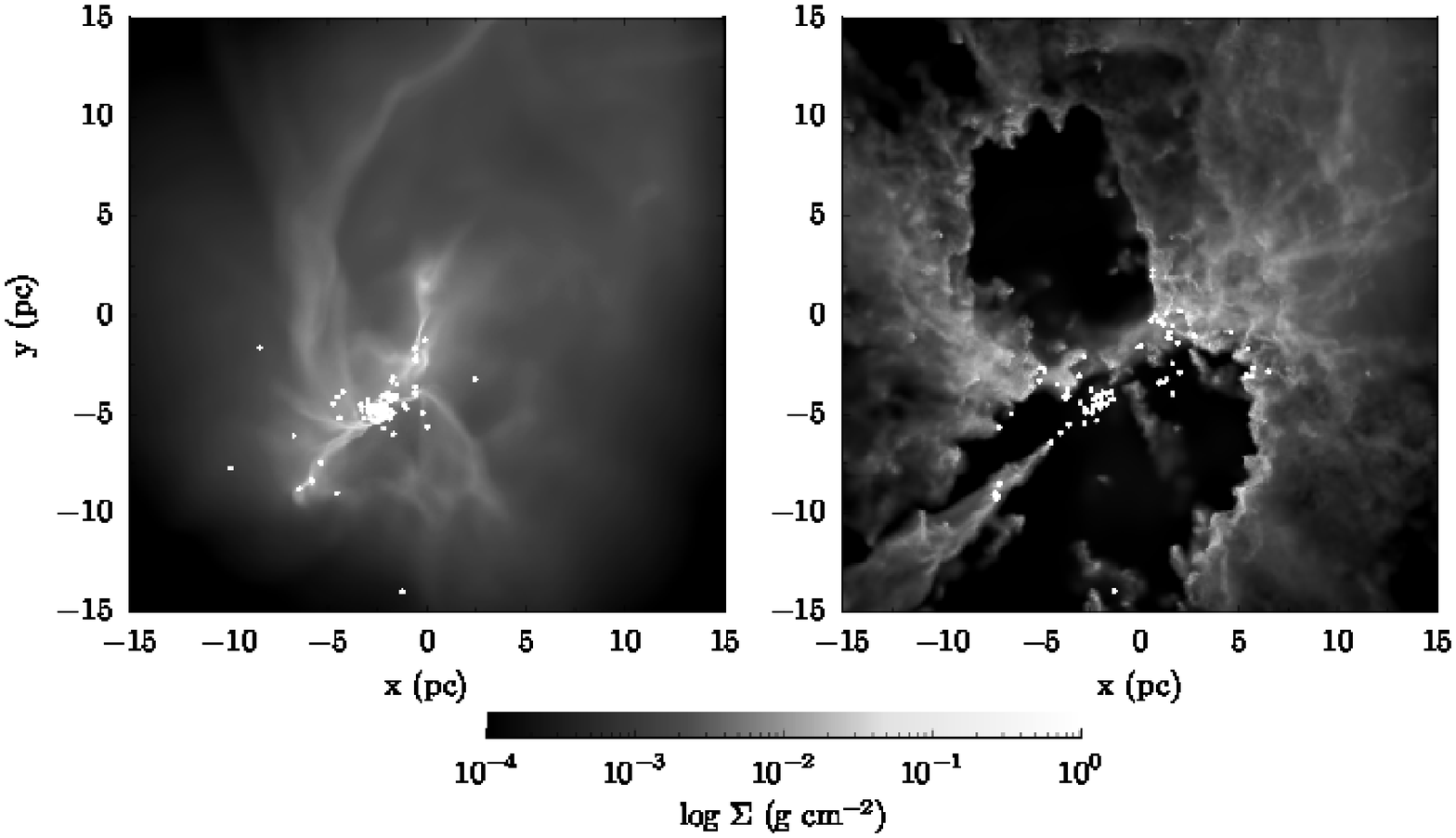}}     
     \hspace{.1in}
     \subfloat[Run J]{\includegraphics[width=0.45\textwidth]{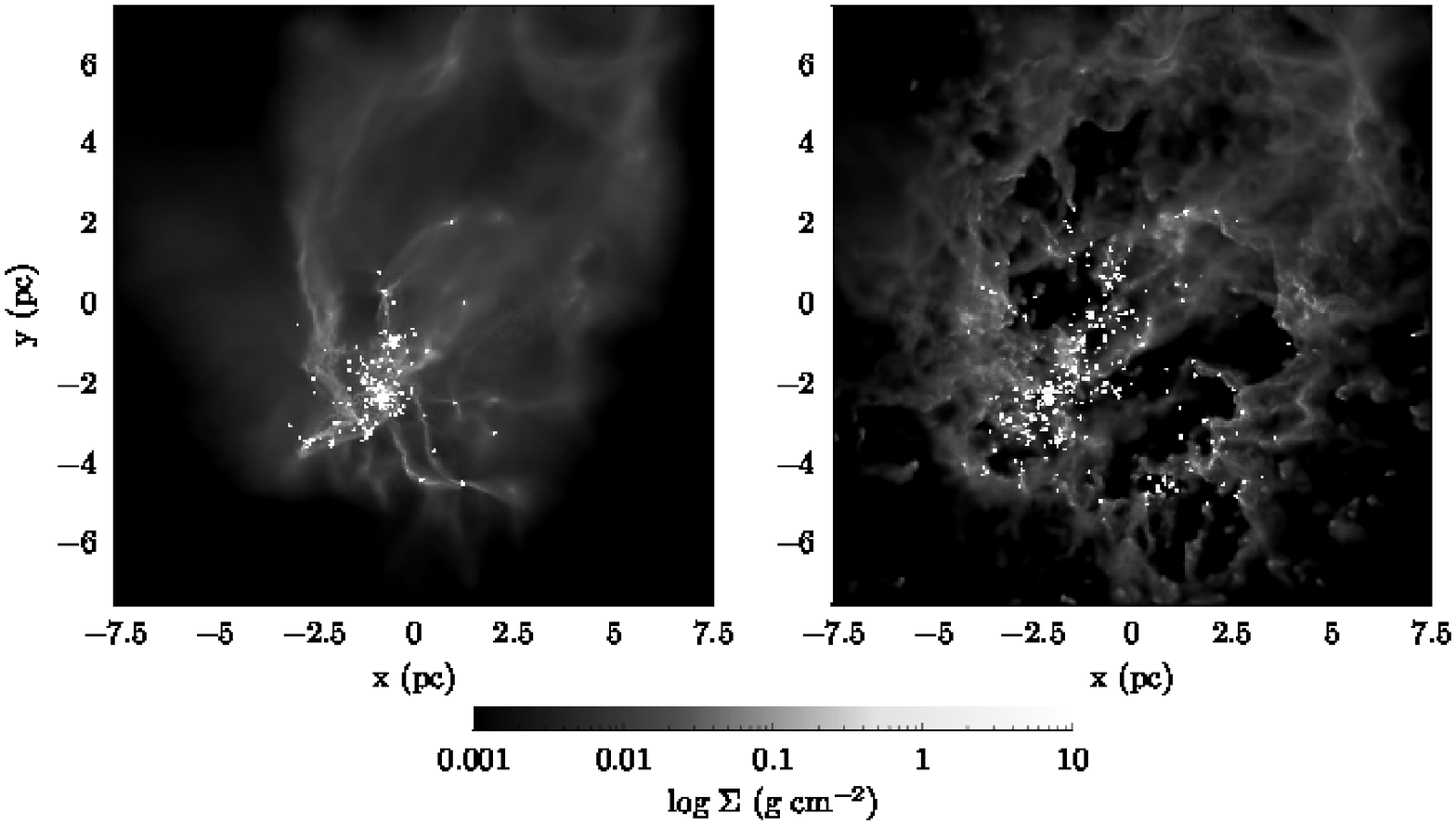}}
     \caption{Comparison of the end states, as viewed down the z--axis, of the control (left panels) and ionized (right panels) Runs A, D, I and J after, respectively, 3.0, 3.0, 2.2 and 1.3 Myr of ionization respectively. Colours are gas column densities and white dots are sink particles representing clusters (Runs A and D) or stars (Runs I and J) and are not shown to scale.}
   \label{fig:compare_end}
\end{figure*}

The impact of ionization on the global appearance of the clouds and their embedded clusters can only be gauged by comparison of the final states of the ionized runs with the corresponding control runs which have evolved in the absence of feedback. In Figure \ref{fig:compare_end}, we compare the end states of the ionized and control runs of clouds A, D, I and J. The principal effect of feedback is, not surprisingly, the creation of bubble--like structures. In the cases of Runs A and J, the interactions of many such bubbles makes the gas morphology in regions affected by feedback extremely complex (although note that large regions of Run A are not influenced by ionization owing to the cloud's large size and sparse star formation. In Runs D and I where ionizing radiation issues largely from a single approximately central cluster, the morphology is simpler, consisting (in projection at least) of a few well--defined bubbles. We also observe champagne flows, where bubbles have burst (in projection) through the borders of the clouds, such as at $\sim(0,-160)$ in Run A, and prominent pillars such as that in the lower left corner of Run I. These structures are, of course, not present in the control runs and thus serve as clear signs of the action of feedback.\\
\subsection{Star formation efficiency and numbers of stars/clusters}
\begin{figure*}
     \centering
     \subfloat[Run A]{\includegraphics[width=0.45\textwidth]{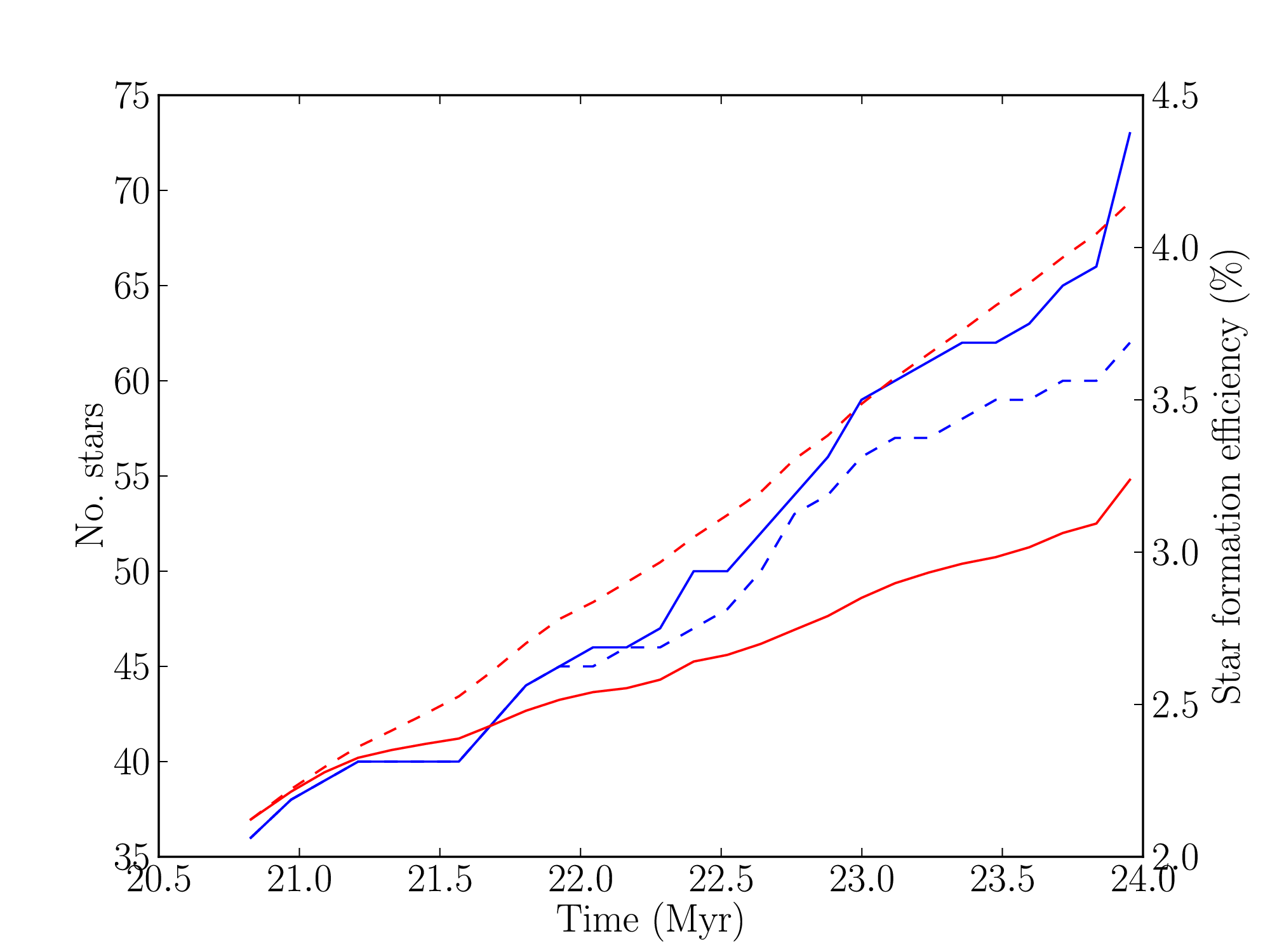}}     
     \hspace{.1in}
     \subfloat[Run D]{\includegraphics[width=0.45\textwidth]{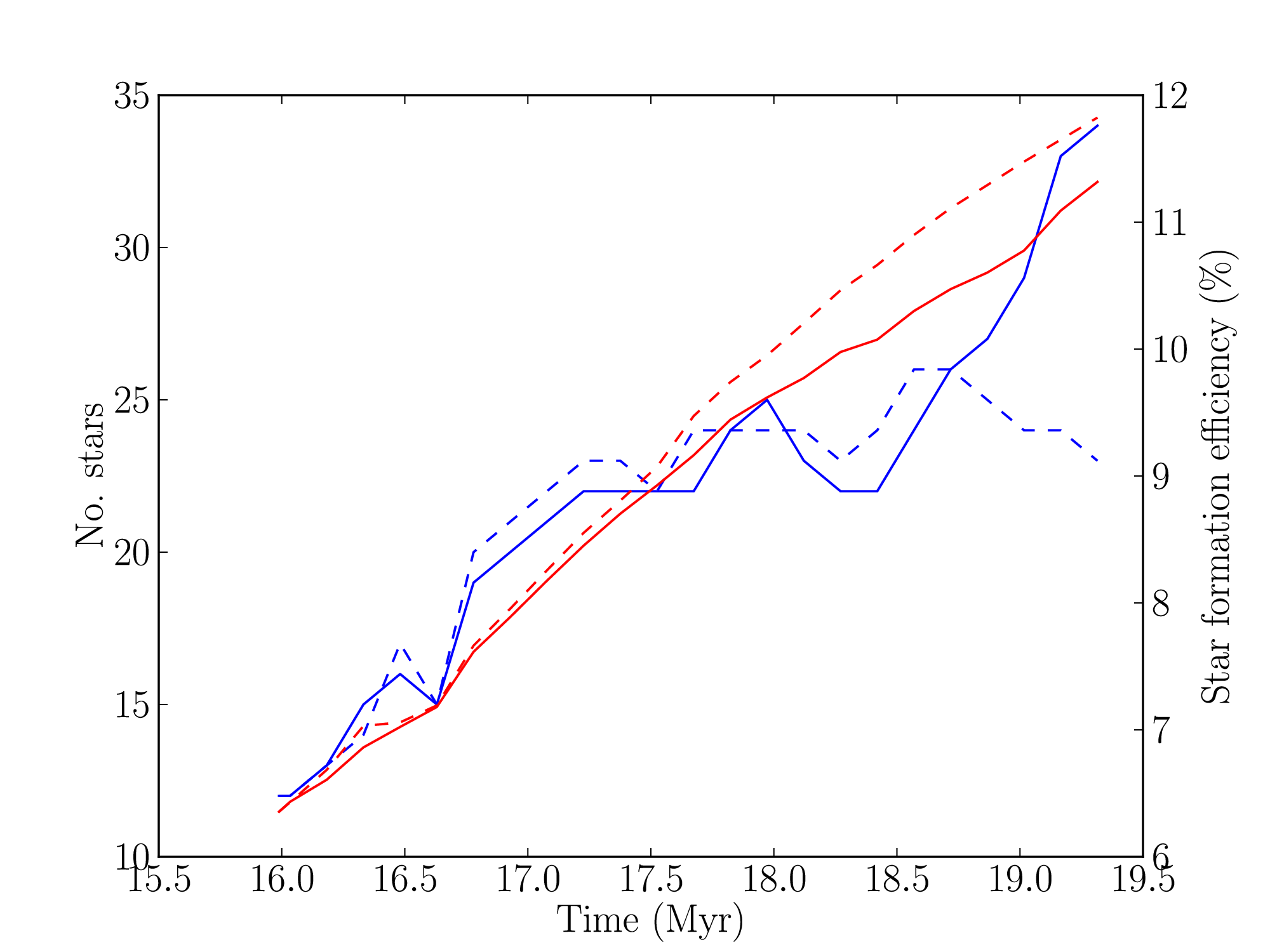}}
     \vspace{.1in}
     \subfloat[Run I]{\includegraphics[width=0.45\textwidth]{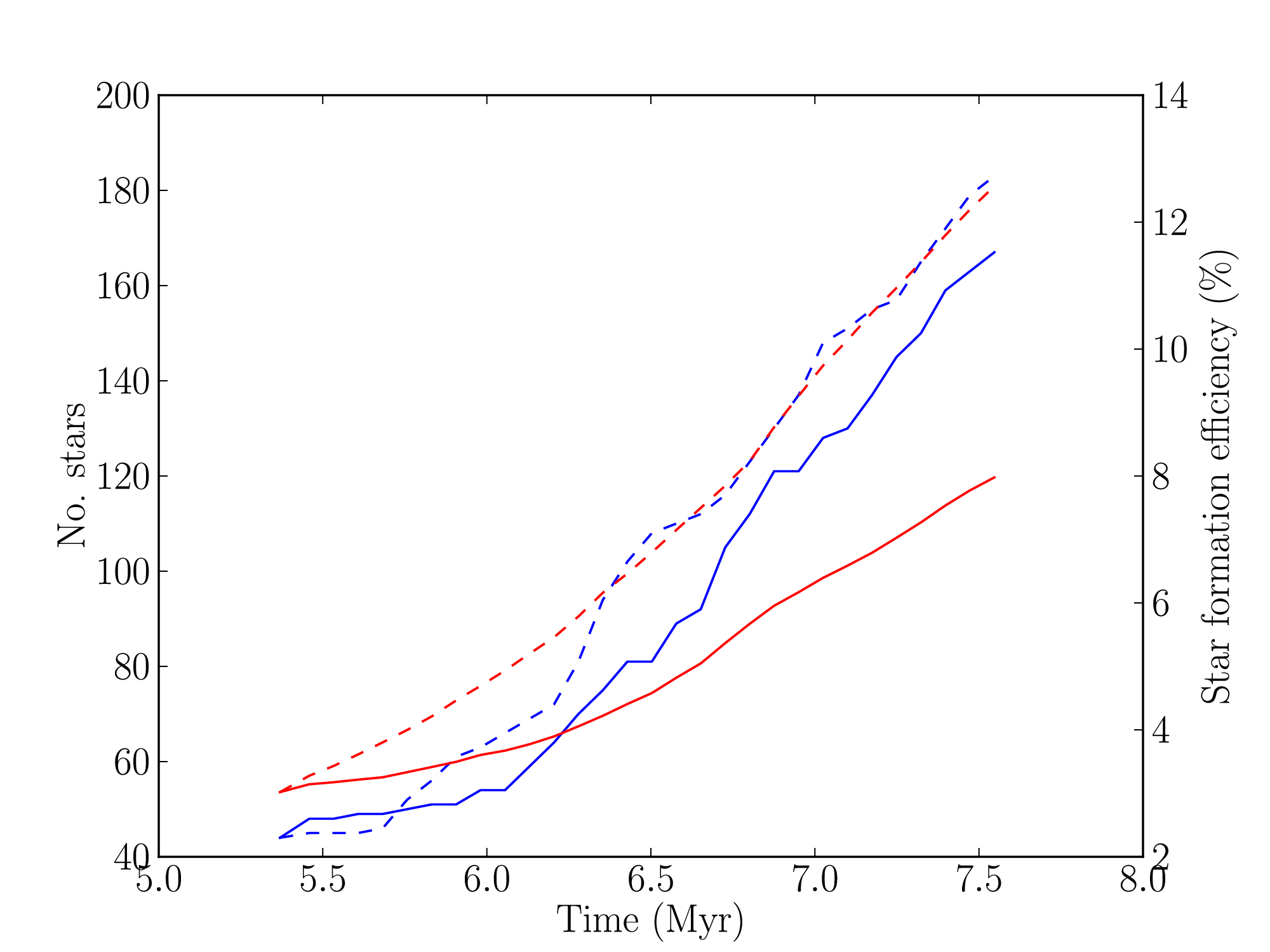}}     
     \hspace{.1in}
     \subfloat[Run J]{\includegraphics[width=0.45\textwidth]{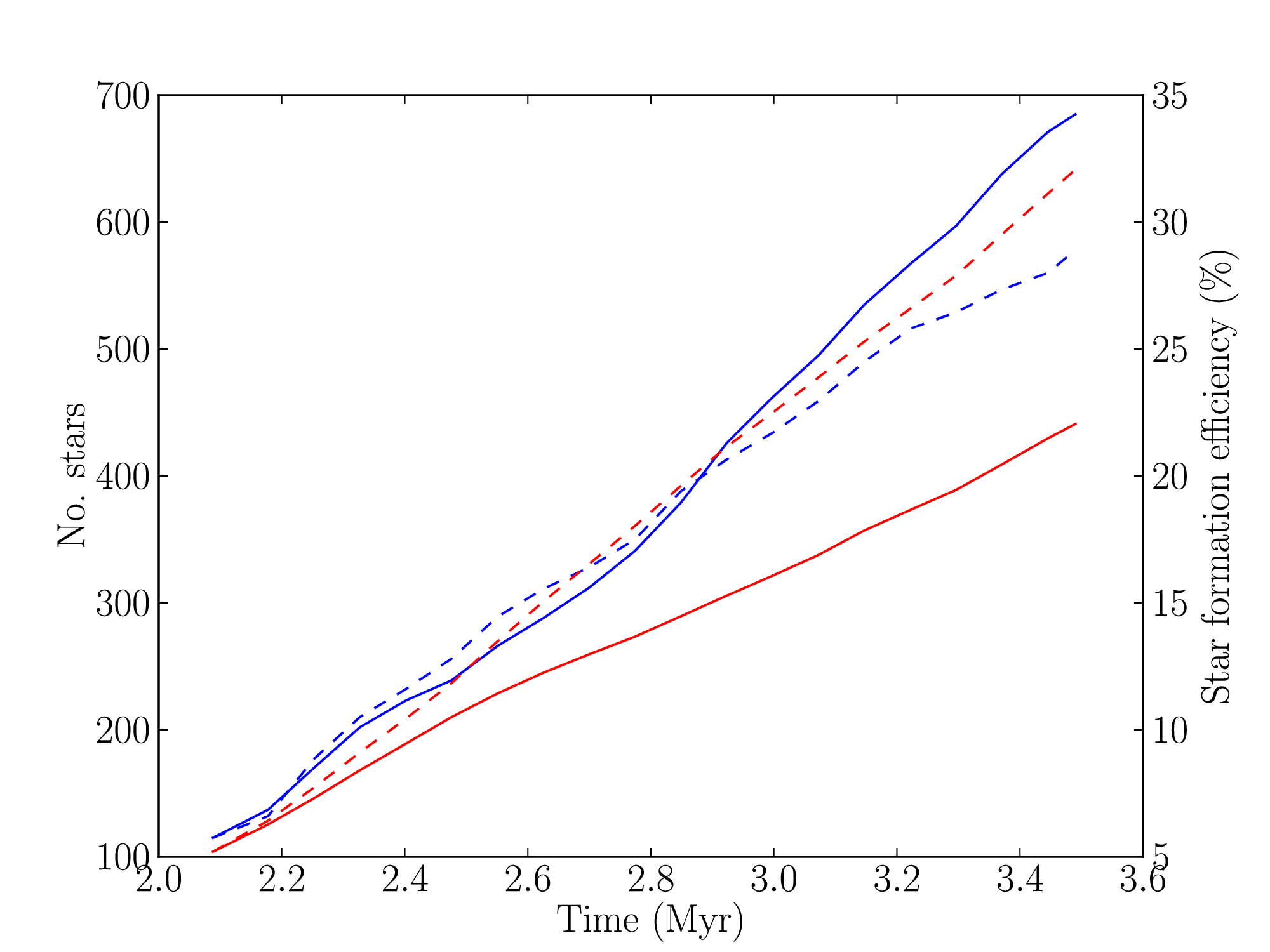}}
     \caption{Comparison of the star--formation efficiency (red lines and right--hand axis) and numbers of stars/clusters (blue lines and left--hand axis) in the ionized (solid lines) and control (dashed lines) Runs A, D, I and J. Star formation efficiencies are lower in all cases in the ionized runs, while numbers of clusters/stars are lower in all ionized runs except Run I.}
   \label{fig:compare_sfe}
\end{figure*}

In Figure \ref{fig:compare_sfe}, we plot against time the star--formation efficiency (red lines) and total numbers of stars or clusters (blue lines) in runs A, D, I and J, with the feedback--affected simulations plotted as solid lines and the control simulations plotted as dashed lines. Note that the reductions in the numbers of clusters in Run D are due to mergers of sink particles (which are also permitted in Run A, but do not actually occur). In all cases, we see that the star--formation efficiency is reduced by the action of feedback, by more than 50$\%$ in the case of Run I, so that the overall effect of feedback on star formation is negative. However, in Runs A, D and J, the numbers of clusters/stars in the feedback runs are significantly larger, whereas in Run I, the feedback run produces slightly \emph{fewer} stars. In Run D, the numbers of mergers occuring in the control and ionized runs are almost identical at 18 and 19 respectively. However, in the control run, the formation of 11 triggered objects, most of which do not experience a merger, partially offsets this effect, resulting in a larger number of objects in the ionized calculation. In general, in all runs, the star--formation efficiency is lower and the number of stars or clusters is not much changed, and often somewhat higher, in the ionized runs, leading to lower average star or cluster masses in the feedback--influenced systems.\\
\indent As was explained in \cite{2007MNRAS.375.1291D} and \cite{2012MNRAS.tmp.2723D}, unequivocally identifying star formation triggered by feedback requires the comparison of counterfactuals, i.e. comparing the evolution of a given system in the presence and absence of feedback. While this is impossible from an observational standpoint, it is relatively straightforward in the context of simulations, particularly in Lagrangian calculations. The hope is that an improved theoretical understanding of triggering will translate to a better understanding of real systems whose evolution cannot be rerun.\\
\indent In order to quantitatively study the prevalence of triggering in the simulations presented in Paper 1, we first selected those simulations on which ionization was observed to have a strong effect, namely A, D, I and J and repeated them from the epoch at which ionization was turned on, but with feedback disabled. We then use techniques similar to those described in \cite{2007MNRAS.375.1291D} and \cite{2012MNRAS.tmp.2723D} to determine which stars in the control runs also form in the ionized runs and vice versa. A sink which forms in a given control run but not in the corresponding ionized run can be thought of as aborted, whereas a star that forms in the ionized run but not the corresponding control run has been triggered.\\
\subsection{Triggering}
\begin{figure*}
     \centering
     \subfloat[Run A]{\includegraphics[width=0.45\textwidth]{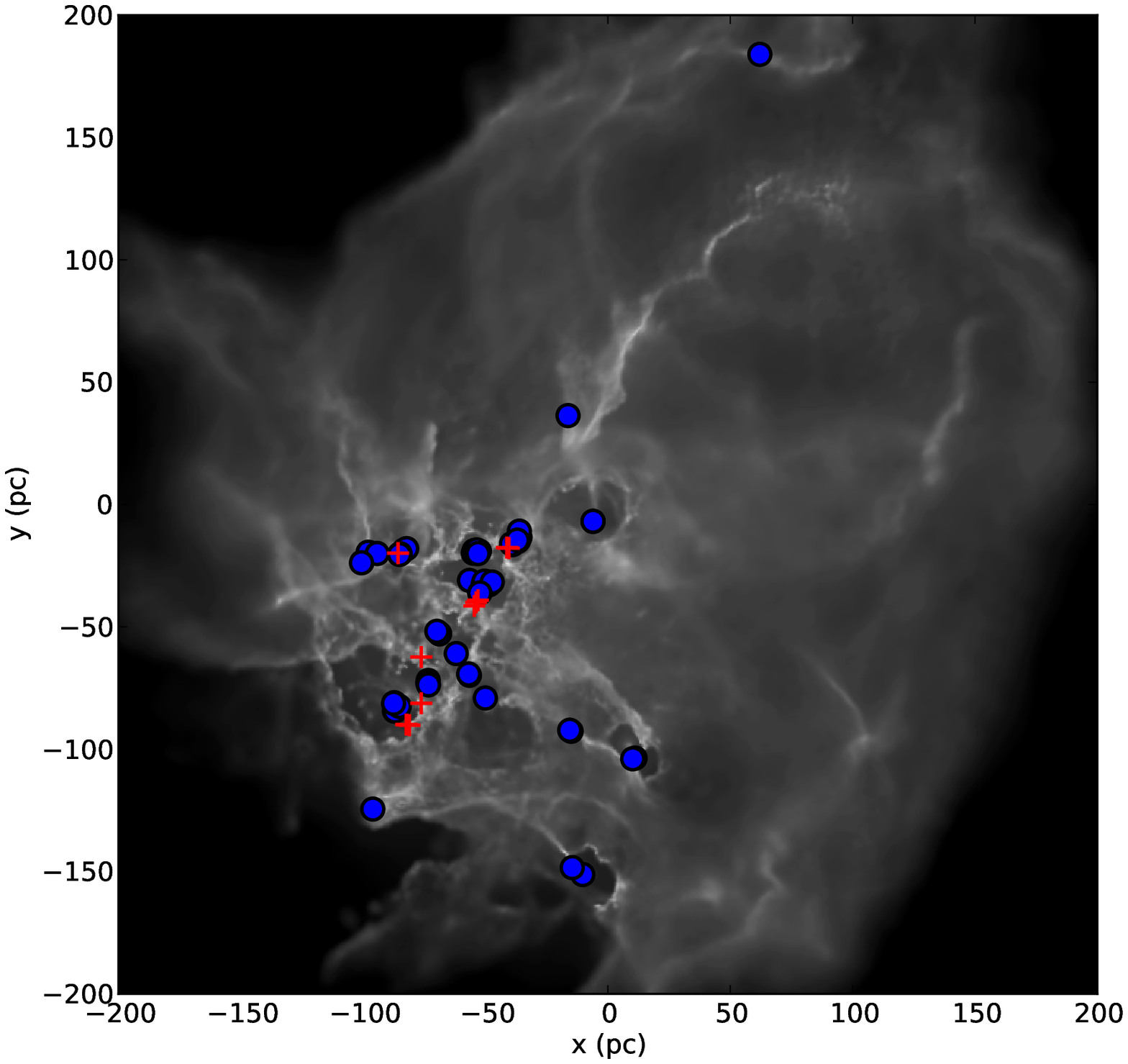}}     
     \hspace{.1in}
     \subfloat[Run D]{\includegraphics[width=0.45\textwidth]{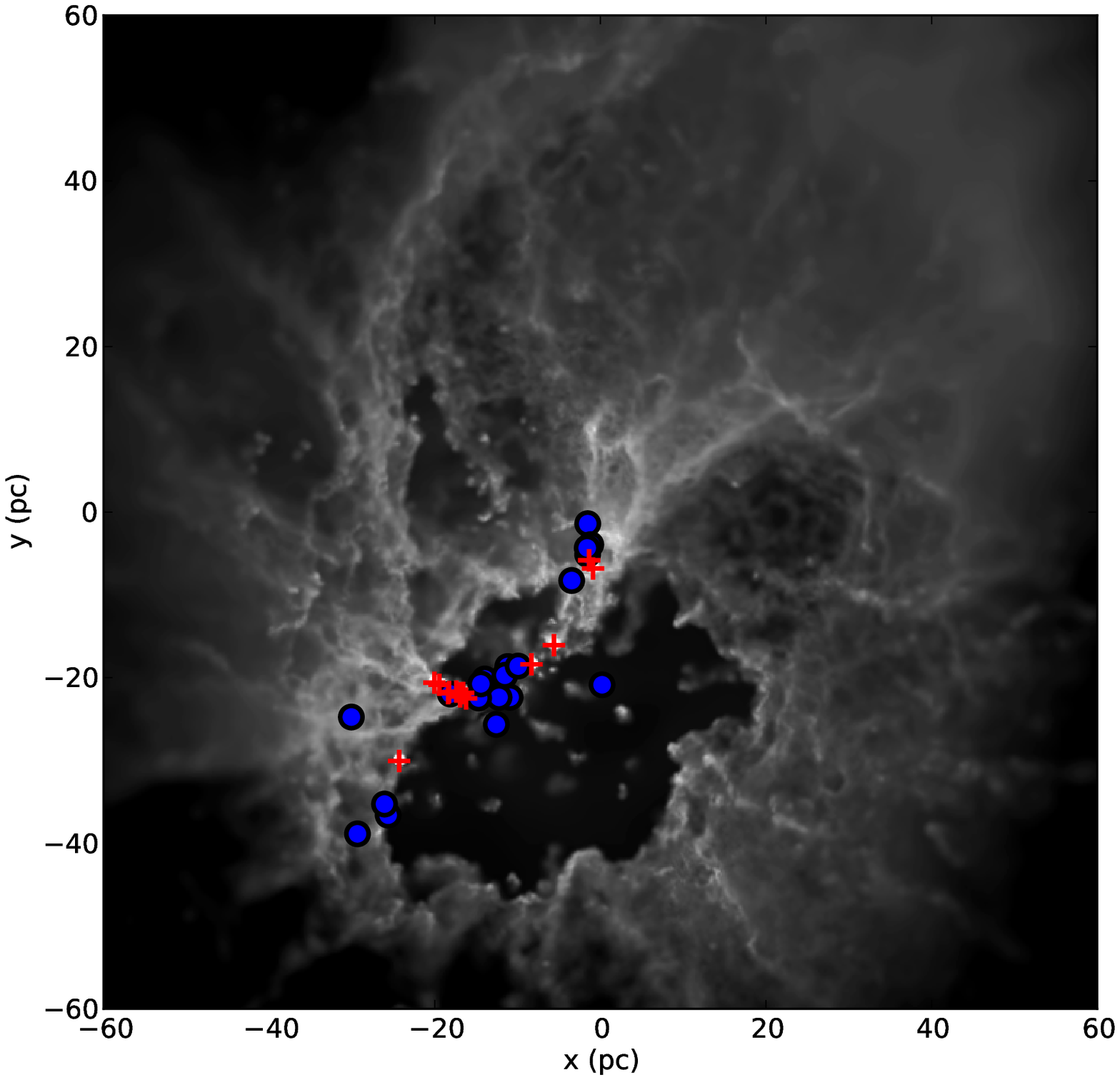}}
     \caption{Locations of triggered (red crosses) and spontaneously--formed (blue circles) clusters compared to the gas column density, shown in greyscale, in the ionized runs A (left) and D (right) as viewed along the z--axis.}
   \label{fig:trigAD}
\end{figure*}

\begin{figure*}
\centering
     \subfloat{\includegraphics[width=0.45\textwidth]{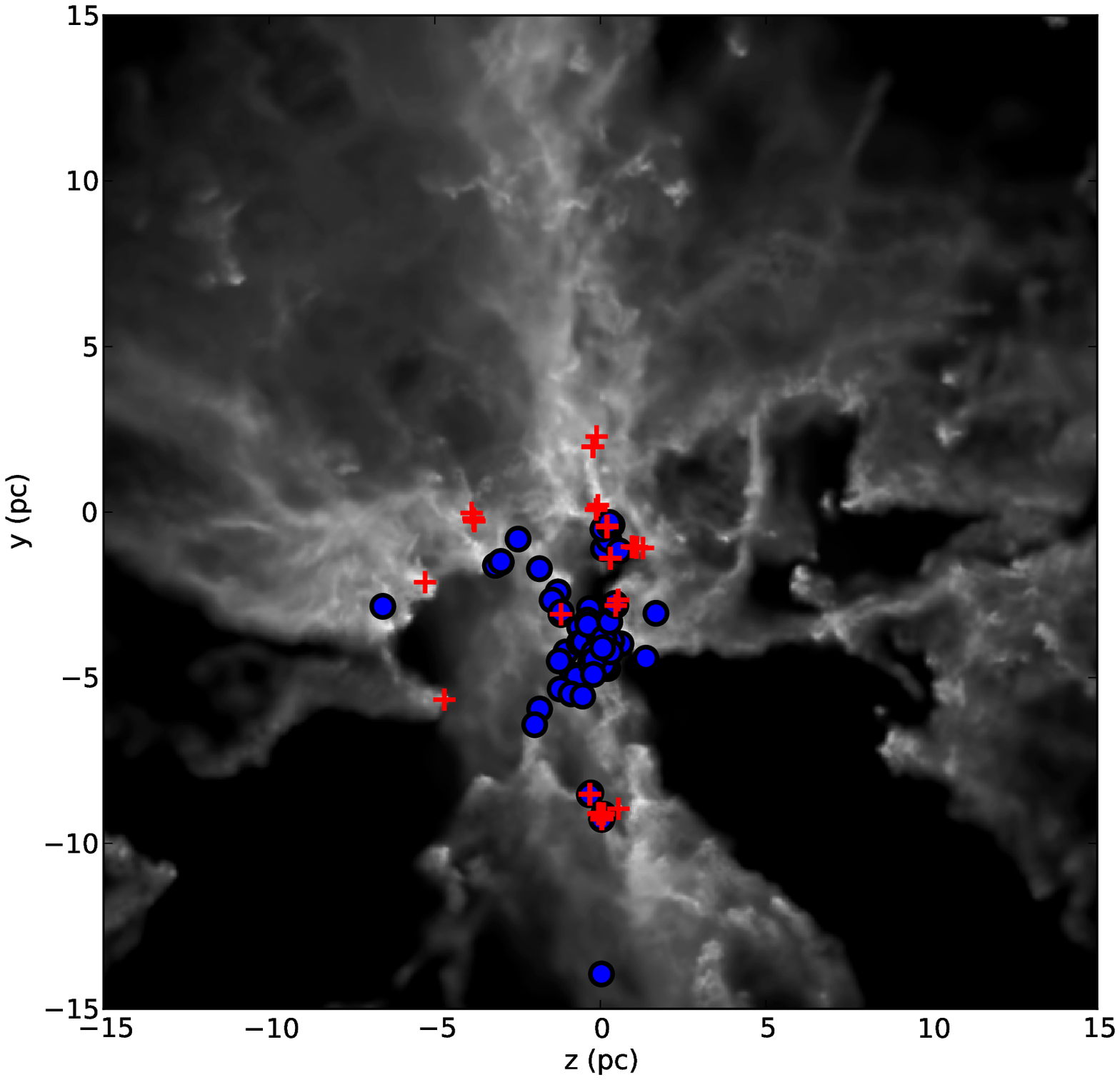}}     
     \hspace{.1in}
     \subfloat{\includegraphics[width=0.45\textwidth]{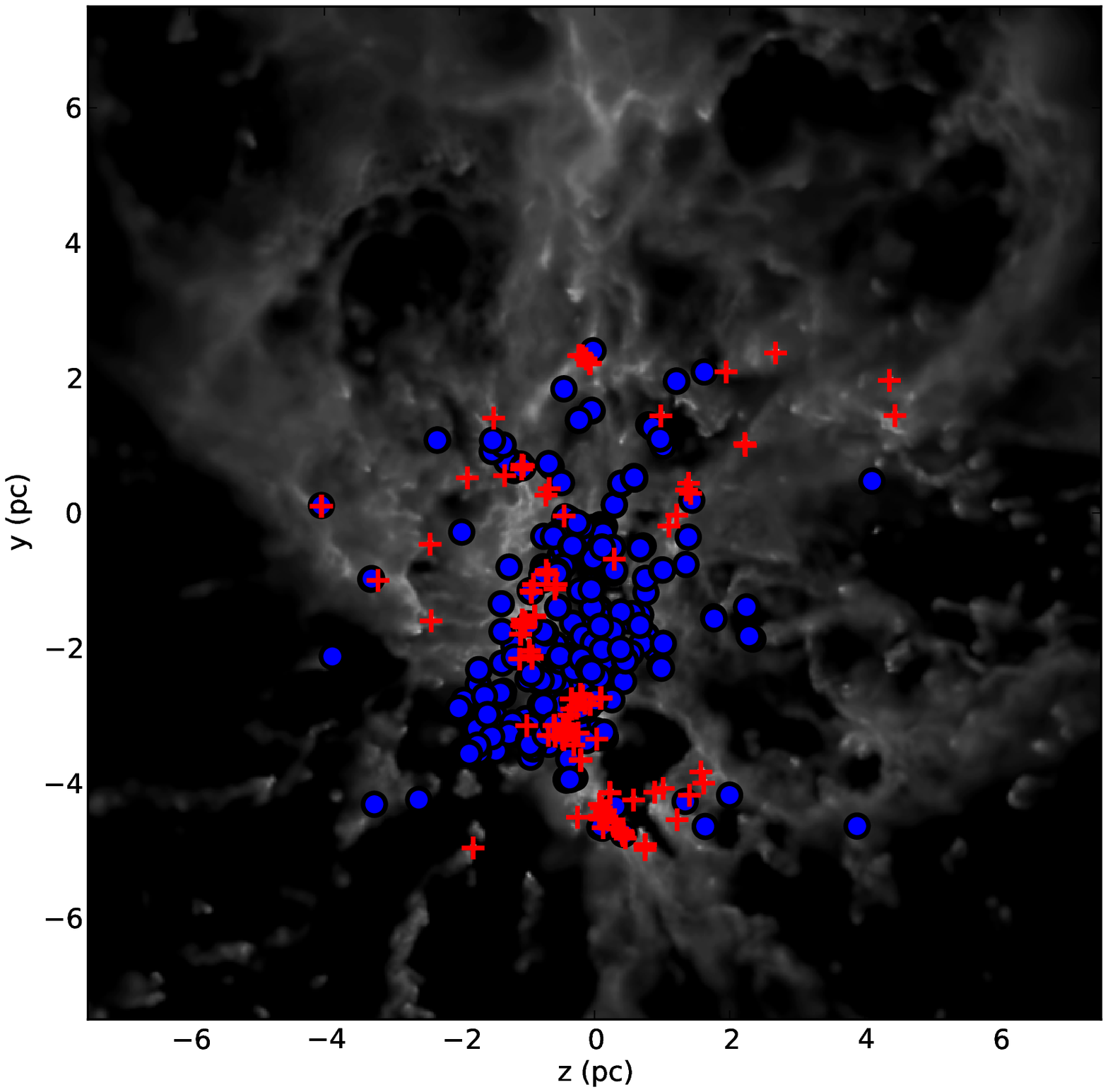}}
     \vspace{-.2in}
     \subfloat{\includegraphics[width=0.45\textwidth]{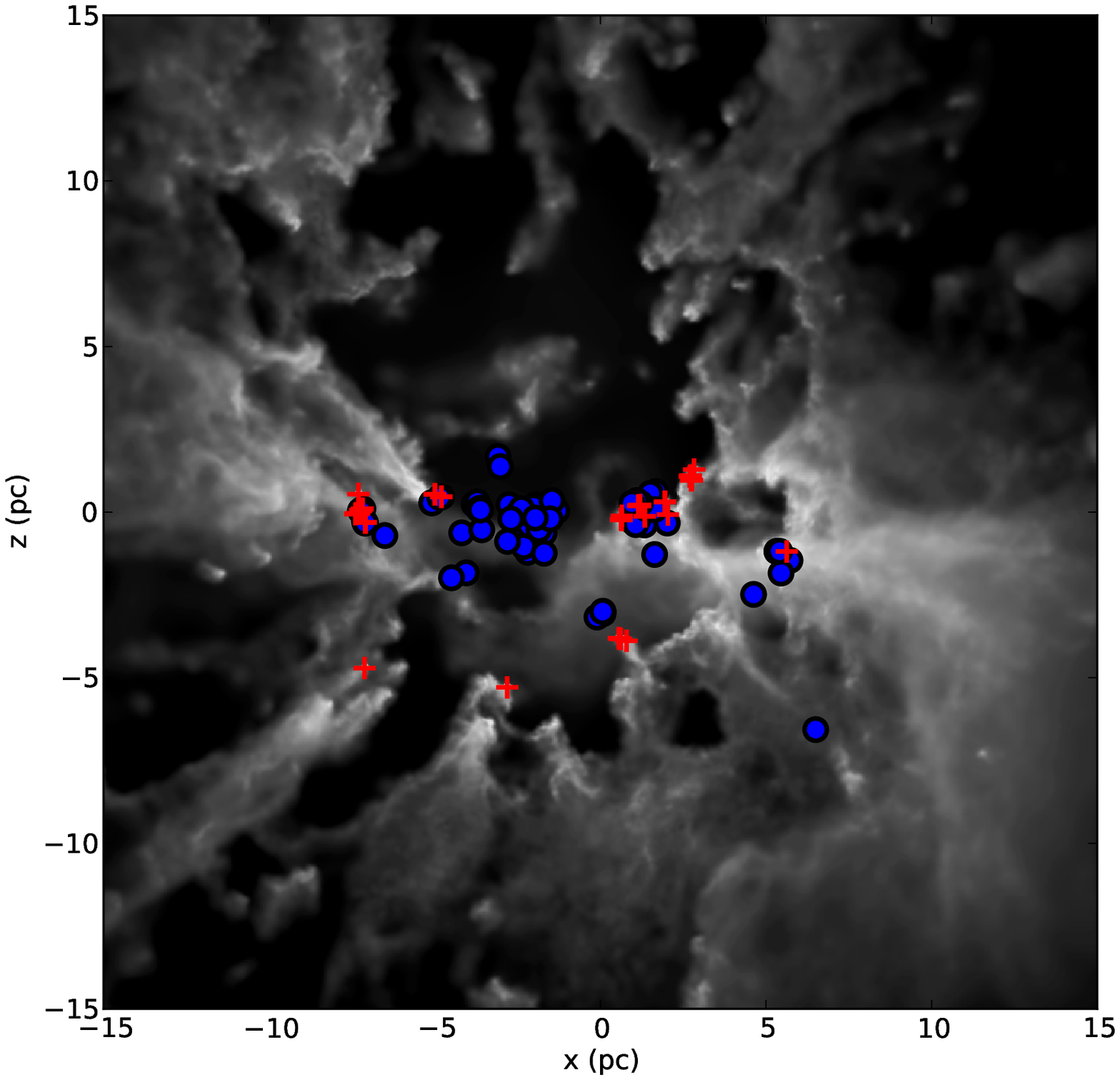}}     
     \hspace{.1in}
     \subfloat{\includegraphics[width=0.45\textwidth]{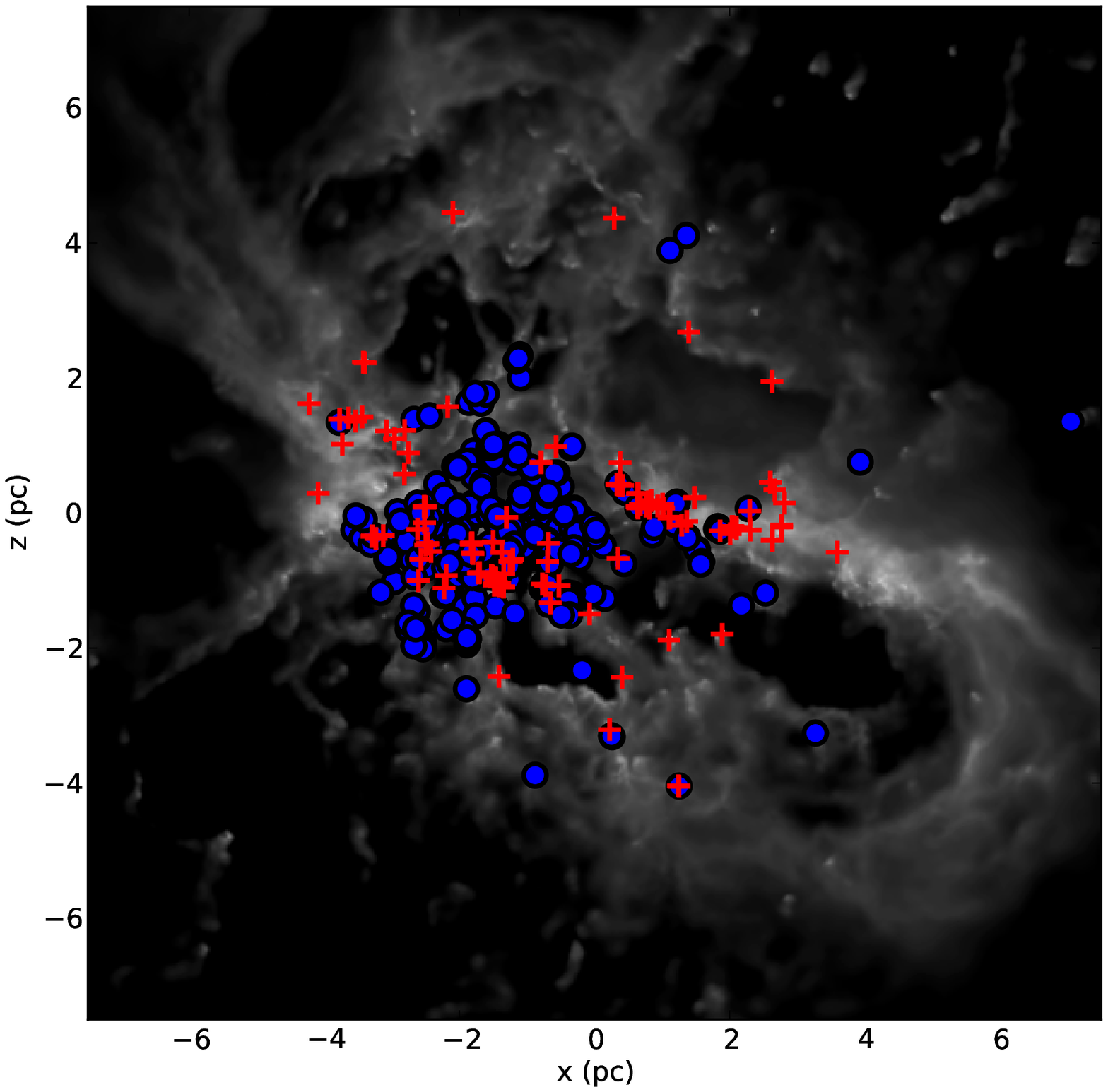}}
     \vspace{-.2in}
     \subfloat[Run I]{\includegraphics[width=0.45\textwidth]{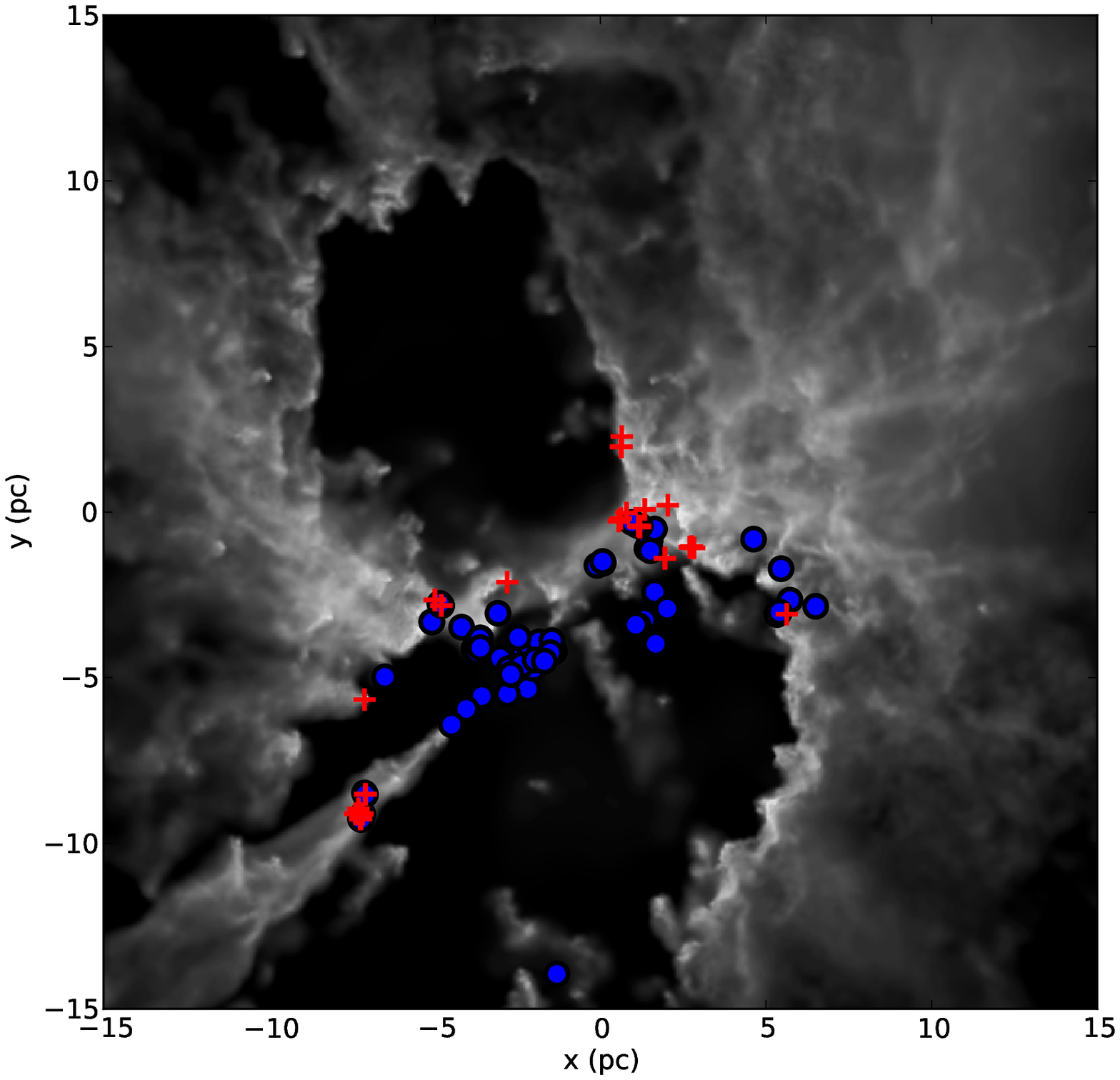}}     
     \hspace{.1in}
     \subfloat[Run J]{\includegraphics[width=0.45\textwidth]{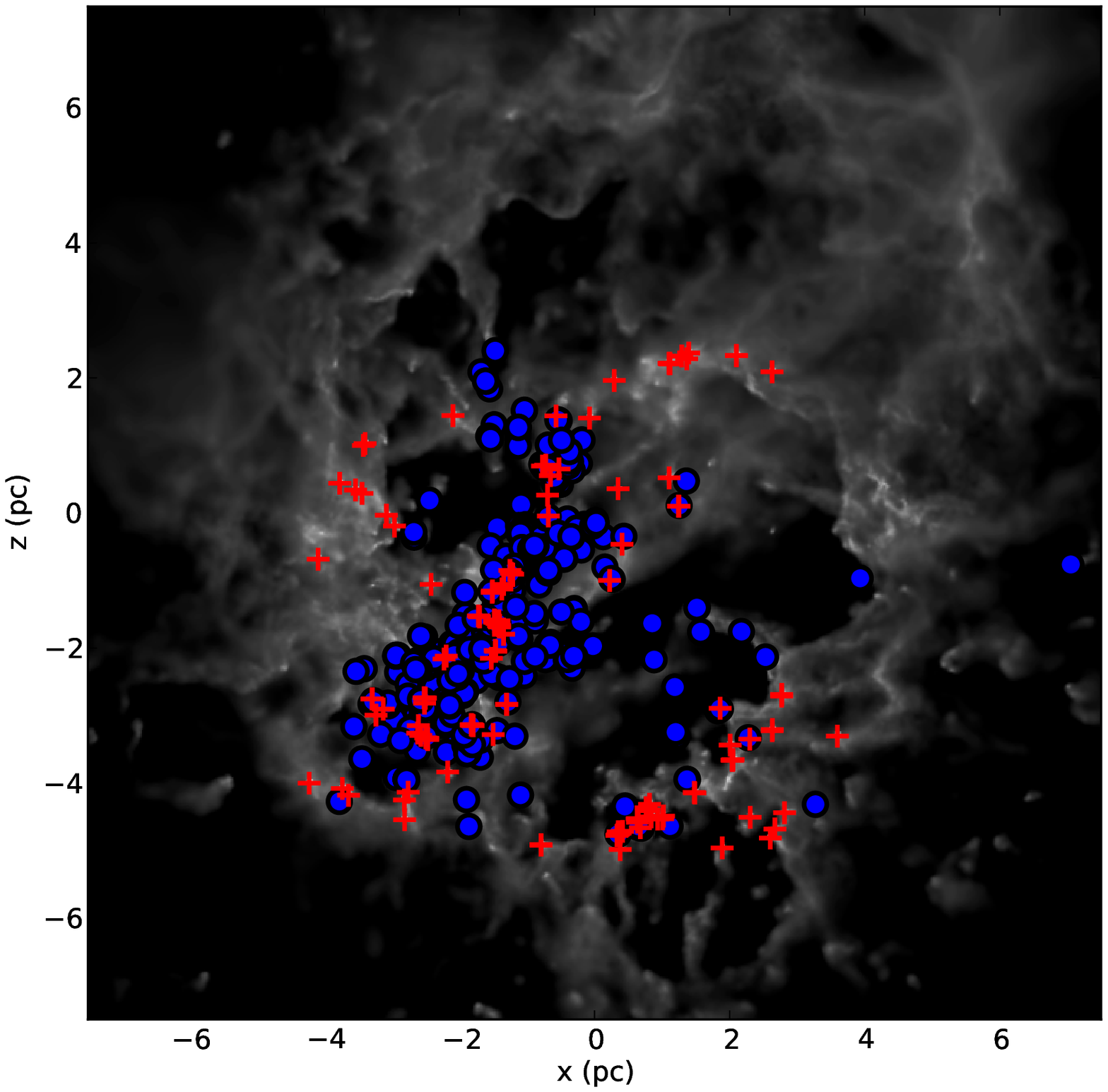}}
\caption{Locations of triggered (red crosses) and spontaneously--formed (blue circles) stars compared to the gas column density, shown in greyscale, in the ionized runs I (left column) and J (right column) as viewed down the x (top row), y (middle row) and z (bottom row) axes).}
   \label{fig:trigIJ}
\end{figure*}

\indent We noted in \cite{2012MNRAS.tmp.2723D} that, in complex systems where star formation is vigorous, it is often the case that the gas from which stars are forming is merely mixed or stirred by feedback, but not necessarily disrupted, expelled or prevented from forming stars. It can therefore be, in principle, that the same gas is converted to stars in both control and ionized simulations, but that it is never the case that the \emph{same star} forms in both simulations. We here consider three ways of comparing stars between simulations: (A) \textit{same seed method} -- here we trace only the $\sim100$ \textit{seed} particles from which each sink particle initially forms (as distinct from those which it accretes later) and see if $\ge50\%$ those same particles seed a single sink particle in the companion run; (B) \textit{same star method} -- this is an extension of method (A) in which we trace all particles from which a sink formed -- seed particles as well as those subsequently accreted -- and determine whether a large fraction ($\ge50\%$) of these particles form a single sink in the companion run; (C) \textit{involved method} -- this traces all particles from which each sink forms and asks only whether $\ge50\%$ of them are involved in star formation in the companion run. Method (A) really traces \emph{star--formation events} -- the initial collapse of a core to form a protostar. Method (B) instead follows the whole process leading to the determination of an object's final mass and is of more interest in general. Both the same--sink and same--star method tie the concept of triggering to individual objects and ask whether the same objects from in the two runs or not, whereas method (C) is less restrictive and concerns the fate of the star--forming gas, determining whether or not stars in the two compared runs are forming from the same pool of material. In Table \ref{tab:trig_analysis}, we present the results of analysing the four simulations using each of these three methods. The columns in the table are: total numbers of stars/clusters in the ionized runs before ionization (N$_{\rm TOT}^{0}$), total numbers of stars at the ends of the ionized (N$_{\rm TOT}^{i}$) and control (N$_{\rm TOT}^{c}$), numbers of untriggered objects in runs derived using the three methods described in the text (N$_{\rm untrig}^{\rm A}$, N$_{\rm untrig}^{\rm B}$, N$_{\rm untrig}^{\rm C}$), the fraction of all gas involved in star formation in the ionized runs which was also involved in star formation in the corresponding control run, and the fraction of all gas invoved in the formation of triggered objects in the ionized runs which was also involved in star formation in the control runs.\\
\begin{table*}
\begin{tabular}{|l|l|l|l|l|l|l|l|l|l|}
Run&N$_{\rm TOT}^{0}$&N$_{\rm TOT}^{i}$& N$_{\rm TOT}^{c}$ &N$_{\rm untrig}^{\rm A}$&N$_{\rm untrig}^{\rm B}$&N$_{\rm untrig}^{\rm C}$&Common gas fraction& Involved triggered mass fraction\\
\hline
A&36&73&62&56&58&62&0.92&0.11\\
\hline
D&12&34&23&14&20&25&0.86&0.28\\
\hline
I&46&168&186&47&53&107&0.69&0.22\\
\hline
J&115&685&578&190&167&541&0.77&0.21\\
\end{tabular}
\caption{Total numbers of stars/clusters in the ionized runs before ionization (N$_{\rm TOT}^{0}$), total numbers of stars at the ends of the ionized (N$_{\rm TOT}^{i}$) and control (N$_{\rm TOT}^{c}$), numbers of untriggered objects in runs derived using the three methods described in the text (N$_{\rm untrig}^{\rm A}$, N$_{\rm untrig}^{\rm B}$, N$_{\rm untrig}^{\rm C}$), the fraction of all gas involved in star formation in the ionized runs which was also involved in star formation in the corresponding control run, and the fraction of all gas invoved in the formation of triggered objects in the ionized runs which was also involved in star formation in the control runs.}
\label{tab:trig_analysis}
\end{table*}
\indent In the simulations presented here, the feedback is internally generated, so that star formation must be well underway before any effects can be seen. One would obviously expect that the seed groups of objects that form \emph{before feedback is enabled} must be the same between control and feedback runs, so that they \textit{at least} must be returned as non--triggered by method (A). Columns 2 and 5 in Table \ref{tab:trig_analysis} confirm that this is the case, in that N$_{\rm untrig}^{\rm A}$ is always larger than N$_{\rm TOT}^{0}$ . In the case of Runs D and I, N$_{\rm untrig}^{\rm A}$ is only slightly larger, indicating that in these clouds, as soon as ionization became active, almost all subsequent star formation events were affected by feedback. In Run A by contrast, 37 additional objects formed after the initiation of feedback of which 20 (54$\%$) formed from the same groups of seed particles in both control and ionized runs. In Run J, 670 objects formed after the onset of feedback, of which 75 (11$\%$) formed from the same seed particles across both simulations. These results indicate that in some regions of the Run J cloud and large fractions of the volume of the Run A cloud, star formation was able to proceed unmolested at least for some time after the birth of massive stars elsewhere in the clouds.\\
\indent N$_{\rm untrig}^{\rm B}$ is generally similar to N$_{\rm untrig}^{\rm A}$, indicating that if a given sink was able to form in both feedback and control simulations, its subsequent accretion history was likely to not be strongly modified by the action of feedback, so that methods (A) and (B) lead to similar conclusions on the degree of triggering. As expected, the systems with the lowest degree of triggering have the highest common gas fractions. We also computed the fraction of all gas involved in the formation of triggered objects that actually was involved in star formation in the corresponding control run, for comparison with the 50$\%$ threshold we used to define whether a given object is triggered. We find that the fraction of material forming triggered stars which is involved in star formation in the corresponding control runs is small, well under a third in Run D, little more than one fifth in Runs I and J and one ninth in Run A, so it is clear that most of the total mass going to form the objects we regard as triggered is not involved in star formation in the absence of feedback.\\
\indent In contrast to the results presented here, in \cite{2012MNRAS.tmp.2723D} we found that there was almost no correlation between the sink--particle seeds in the control and feedback simulations, so that the same star--formation event almost never occurred twice. We also found that the correlation between sink--particles accounting for all the mass (method (B)) was also weak, with the same sink particle forming in the feedback and control simulations only $\sim20\%$ of the time. In the simulations presented in that paper, the external radiation source was switched on before any star formation had occurred in the cloud, allowing the shocks driven by the photoevaporation flow to stir the gas for a long period of time before any material became gravitationally unstable and thus making it quite improbable that the same parcel of gas would initiate gravitational collapse or that the same group of gas particles would be subsequently accreted.\\
\indent Method C is the most conservative of the three means of detecting triggering, in that it reports the smallest numbers of triggered or aborted objects and we use the results generated by this method in further discussions. In Figure \ref{fig:trigAD}, we show column--density images of the ionized Runs A and D viewed along the z--axis with the positions of triggered and spontaneously--formed clusters marked as blue and green crosses respectively. Run A exhibits very little triggering while Run D exhibits rather more and the triggered objects are to be found on the peripheries of the ionization--blown bubbles and associated with the dense pillar--like object pointing towards the central concentration of clusters.\\
\indent For Runs I and J, where we resolve individual stars, we plot the systems as viewed along all three principal axes, shown in Figure \ref{fig:trigIJ}. The apparent gas morphology changes quite markedly in Run I, and to a lesser extent in Run J, depending on the viewing angle. Viewed along the z--axis, Run I has a relatively simple form of two large bubbles bisected by a surviving filament of gas, with a large pillar structure (formerly an accretion flow feeding gas into the central cluster) projecting into the lowermost bubble. The bubbles, filament and pillar all appear to be associated with triggered stars. In the y--projection, the bubble morphology is much less pronounced and there are many less well--defined pillars, several of which have associated triggered stars. However, viewed along the x--axis, Run I is much more difficult to interpret. The morphology appears to be a single filament (in fact a sheet seen nearly edge--on) with a small bubble below and left of centre being the only obvious sign of feedback. The stars, triggered or otherwise, are mostly distributed along the filament.\\
\indent The morphology in Run J is rather more complex owing to the greater number and wider distribution of ionizing stars in Run J. This results in a disordered overlapping collection of bubble--like structures as seen down all axes. There are no bubbles as well--defined as those seen in some projections of Run I, nor are there any such obvious pillar structures. The distribution of stars has the approximate form of a centrally--condensed cluster of mostly non--triggered stars, surrounded by a shell of triggered objects embedded within the walls of evacuated cavities.\\
\subsection{Mass functions}
\begin{figure*}
     \centering
     \subfloat[Run A]{\includegraphics[width=0.45\textwidth]{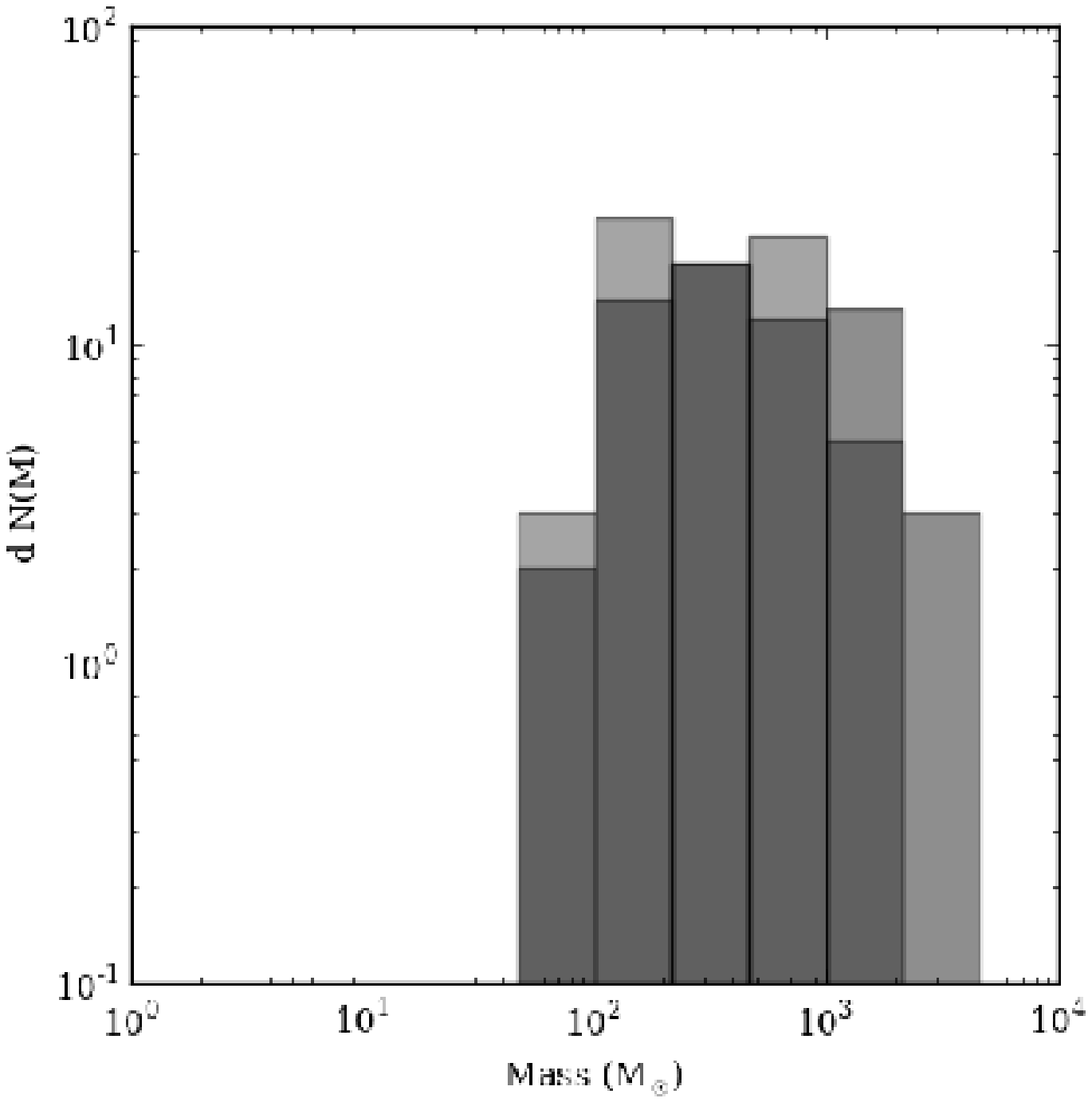}}     
     \hspace{.1in}
     \subfloat[Run D]{\includegraphics[width=0.45\textwidth]{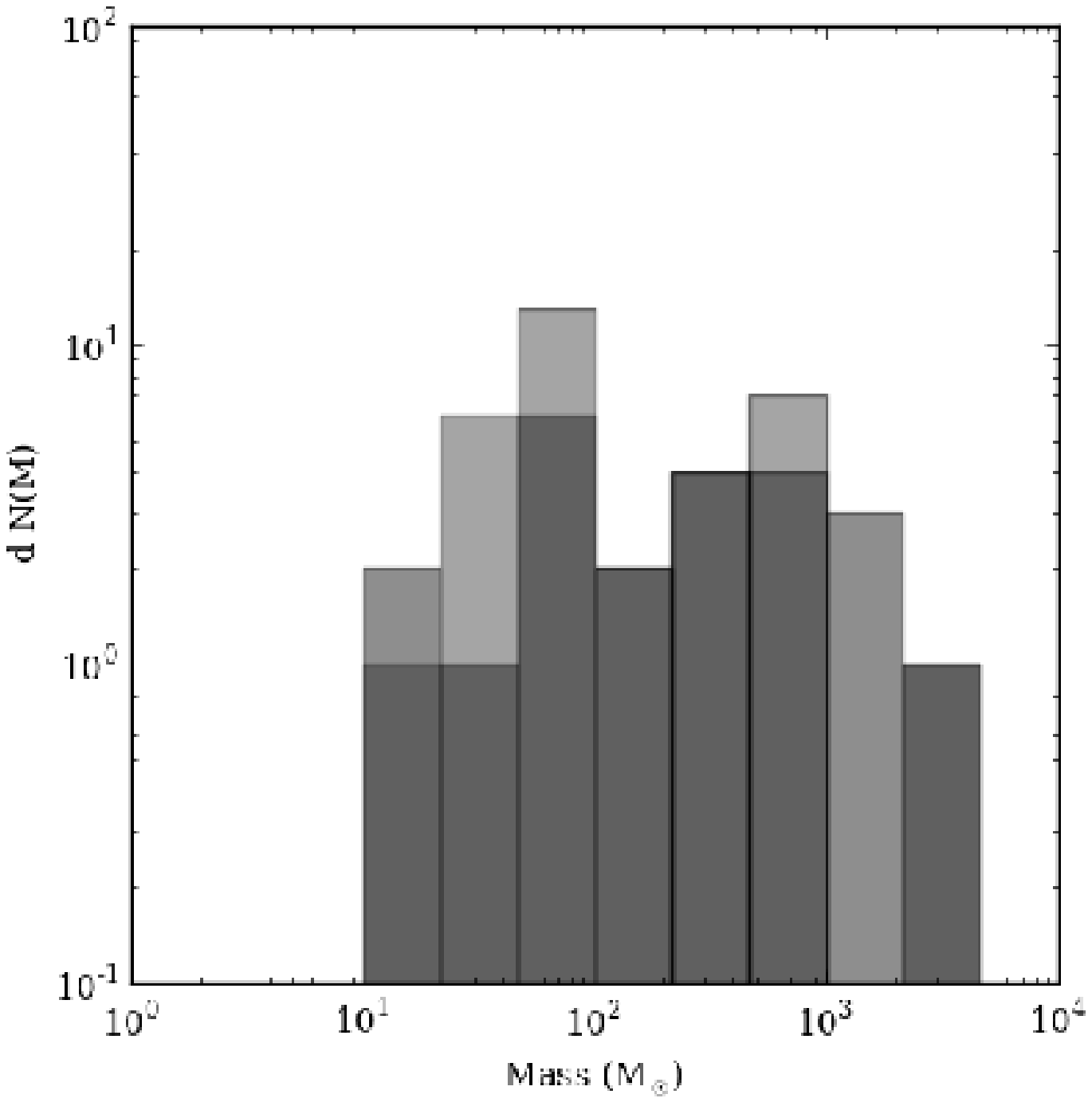}}
     \vspace{.1in}
     \subfloat[Run I]{\includegraphics[width=0.45\textwidth]{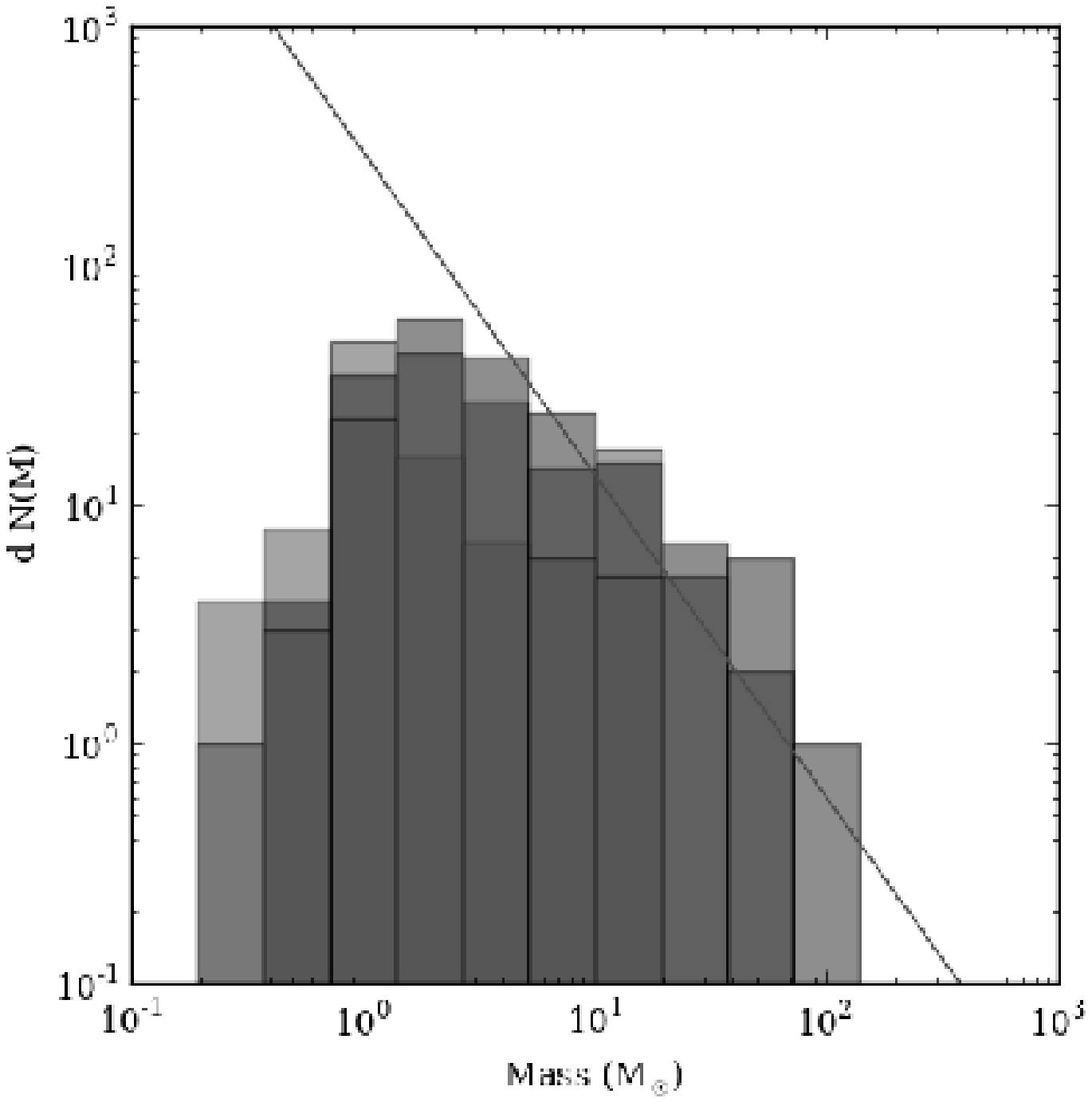}}     
     \hspace{.1in}
     \subfloat[Run J]{\includegraphics[width=0.45\textwidth]{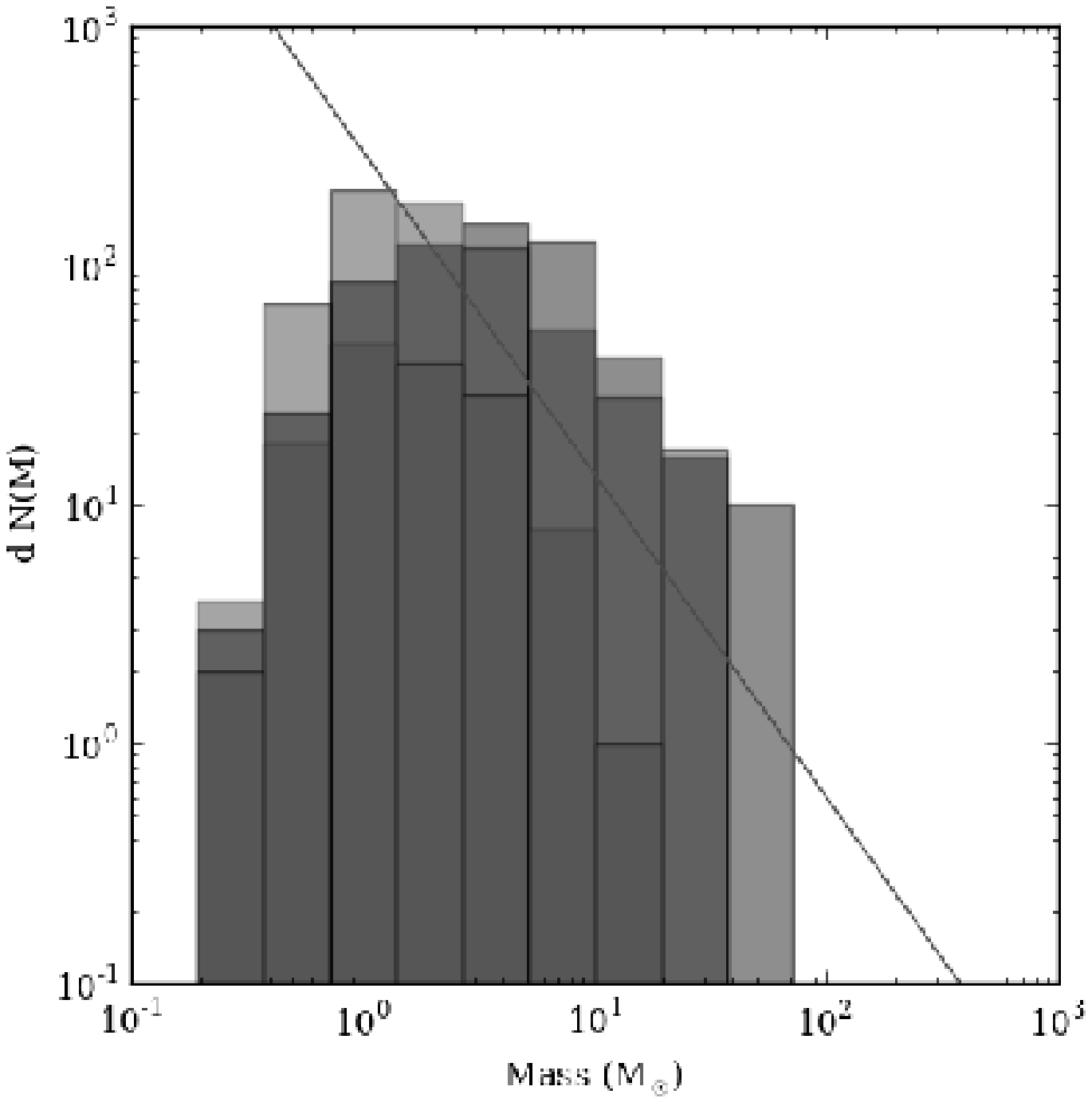}}
     \caption{Comparison of the stellar or cluster mass functions in the ionized (green histograms) and control (blue histograms) Runs A, D, I and J. For Runs I and J, the contribution of triggered objects to the mass function in the ionized run is overlaid in red and the red line denotes the Salpeter slope.}
   \label{fig:compare_massfunc}
\end{figure*}

\begin{figure*}
     \centering
     \subfloat[Run I]{\includegraphics[width=0.45\textwidth]{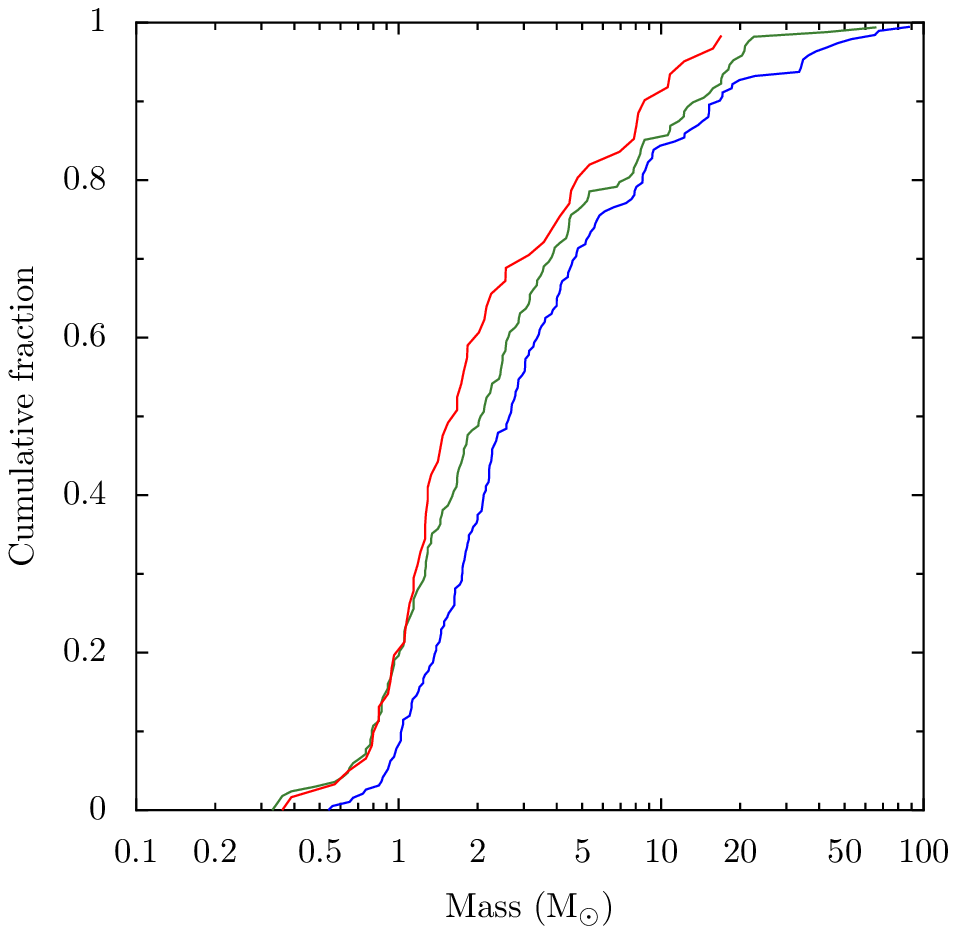}}     
     \hspace{.01in}
     \subfloat[Run J]{\includegraphics[width=0.45\textwidth]{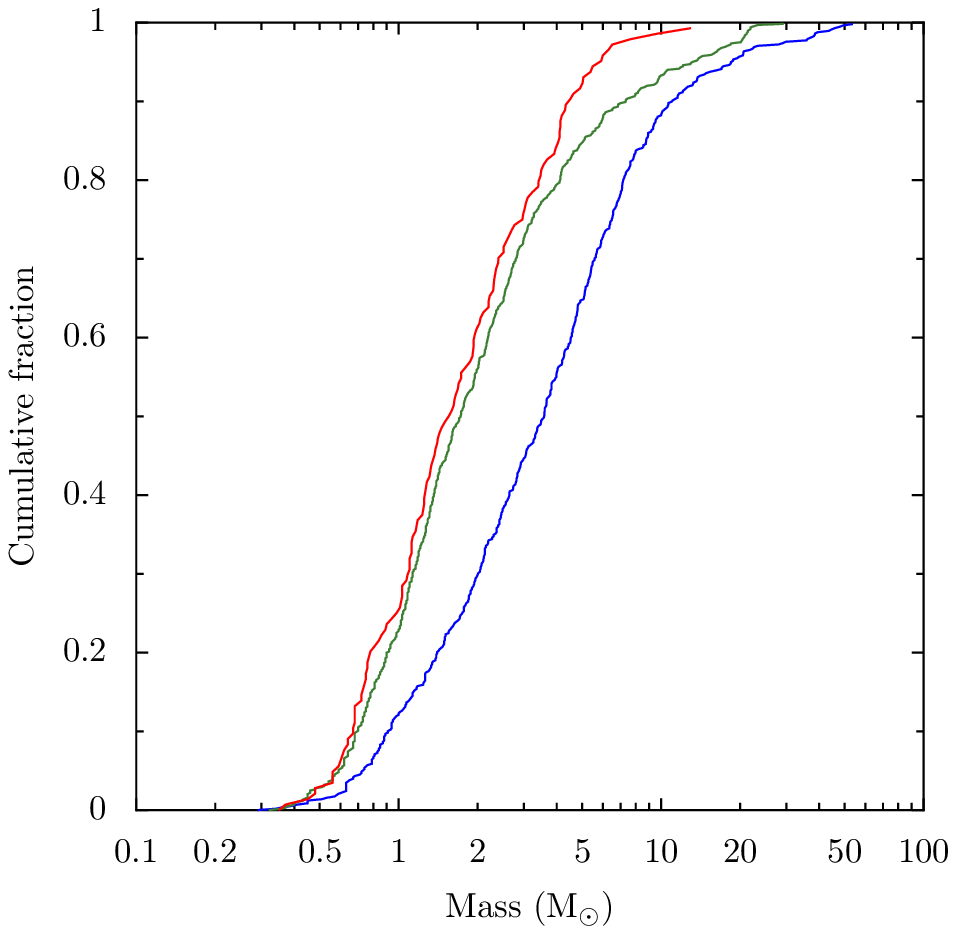}}
     \caption{Comparison of the cumulative stellar mass functions in the feedback (green lines) and control (blue lines) I and J simulations, with cumulative mass functions of triggered objects shown in red. KS tests reveal that the mass functions in the ionized and control runs are significantly different, but that the mass functions of triggered objects are statistically indistinguishable from the total mass functions in the ionized runs, whether objects larger than 20 M$_{\odot}$ are included or not.}
   \label{fig:KS_runij}
\end{figure*}

\begin{figure*}
     \centering
     \subfloat[Run I control]{\includegraphics[width=0.34\textwidth]{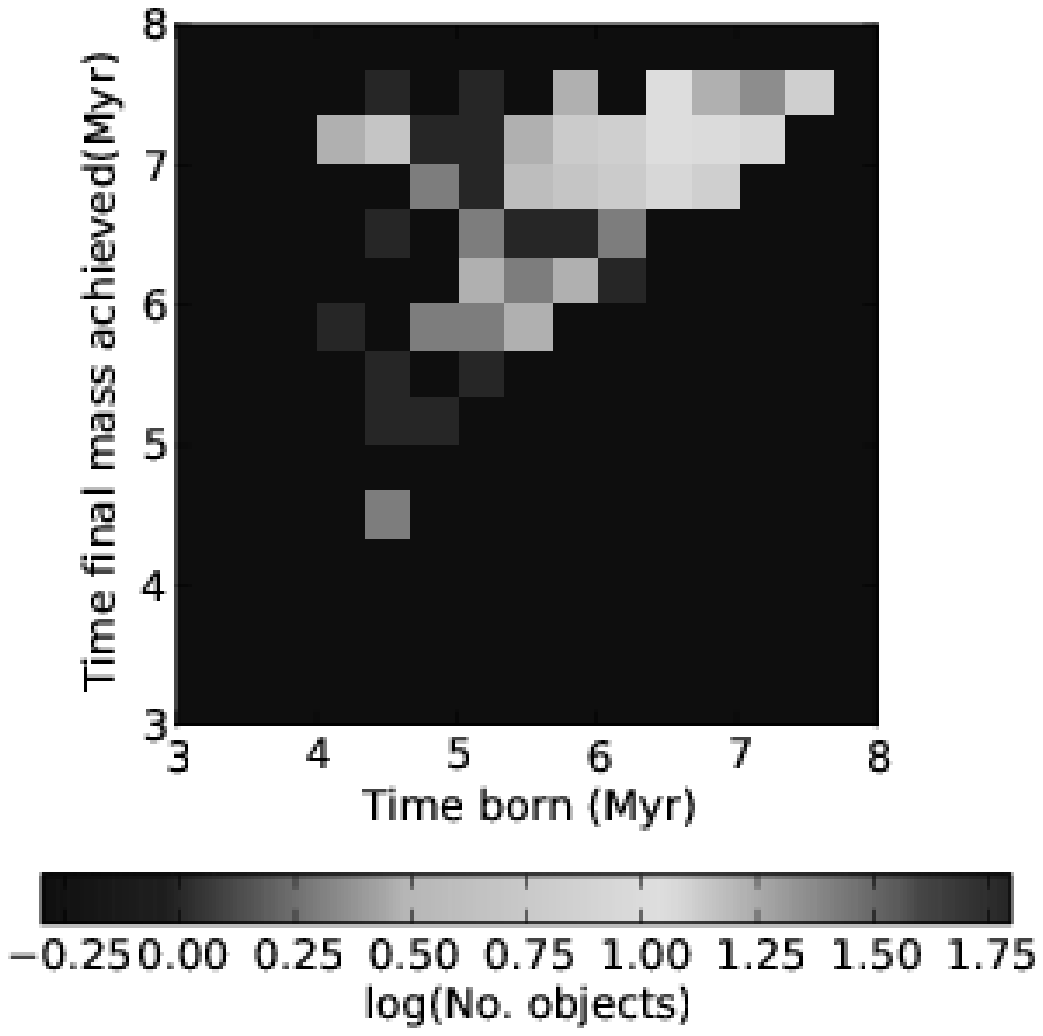}}     
     \hspace{.01in}
     \subfloat[Run I control]{\includegraphics[width=0.34\textwidth]{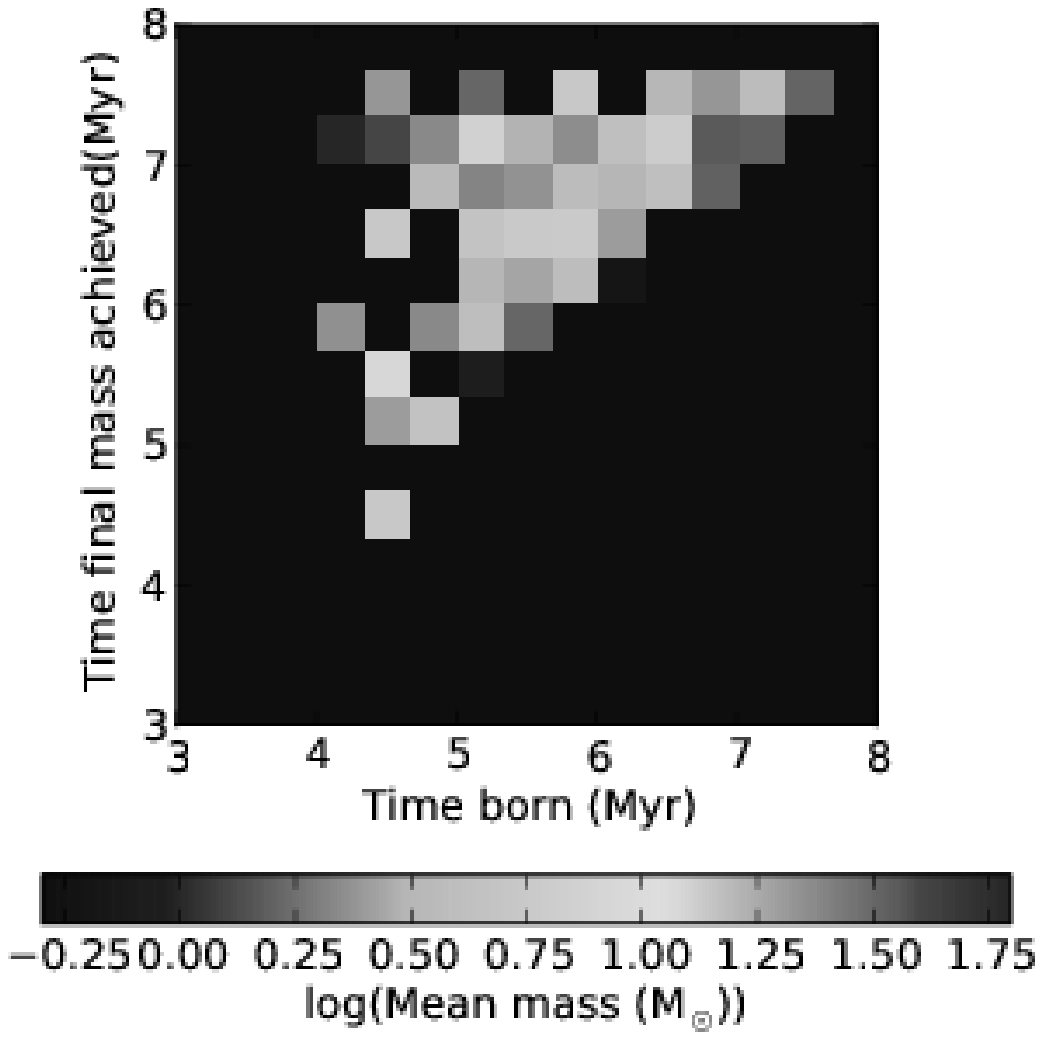}}
     \vspace{.01in}
     \subfloat[Run I ionized]{\includegraphics[width=0.32\textwidth]{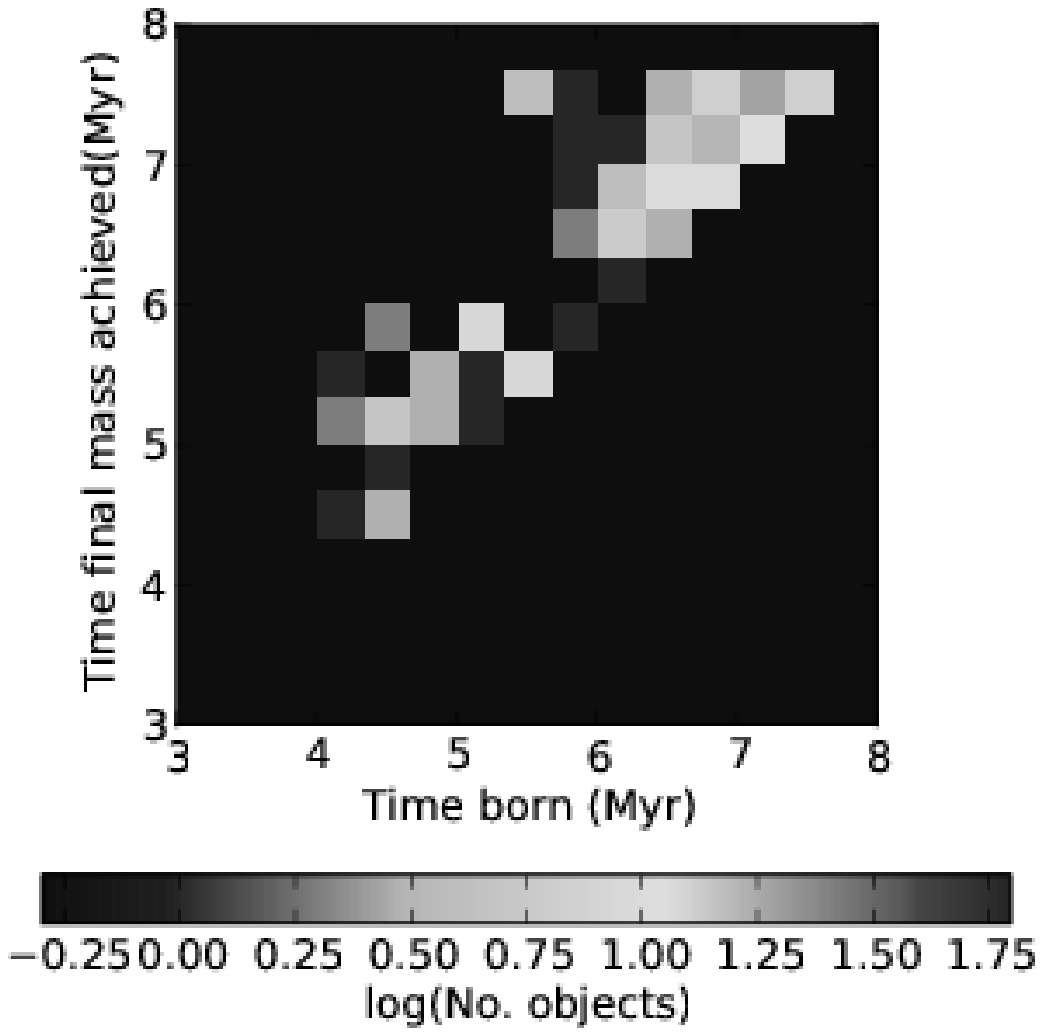}}     
     \hspace{.01in}
     \subfloat[Run I ionized]{\includegraphics[width=0.32\textwidth]{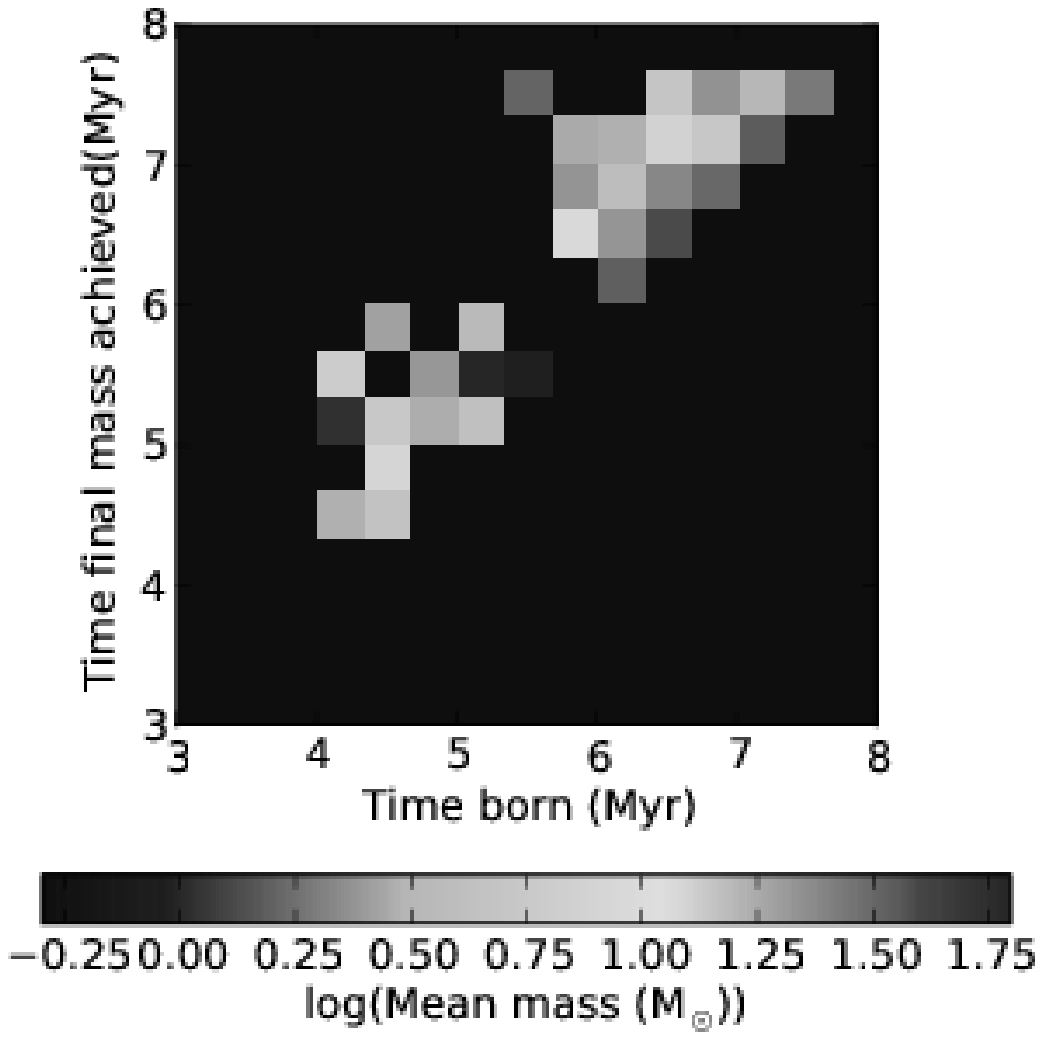}}
    \hspace{.01in}
     \subfloat[Run I ionized]{\includegraphics[width=0.32\textwidth]{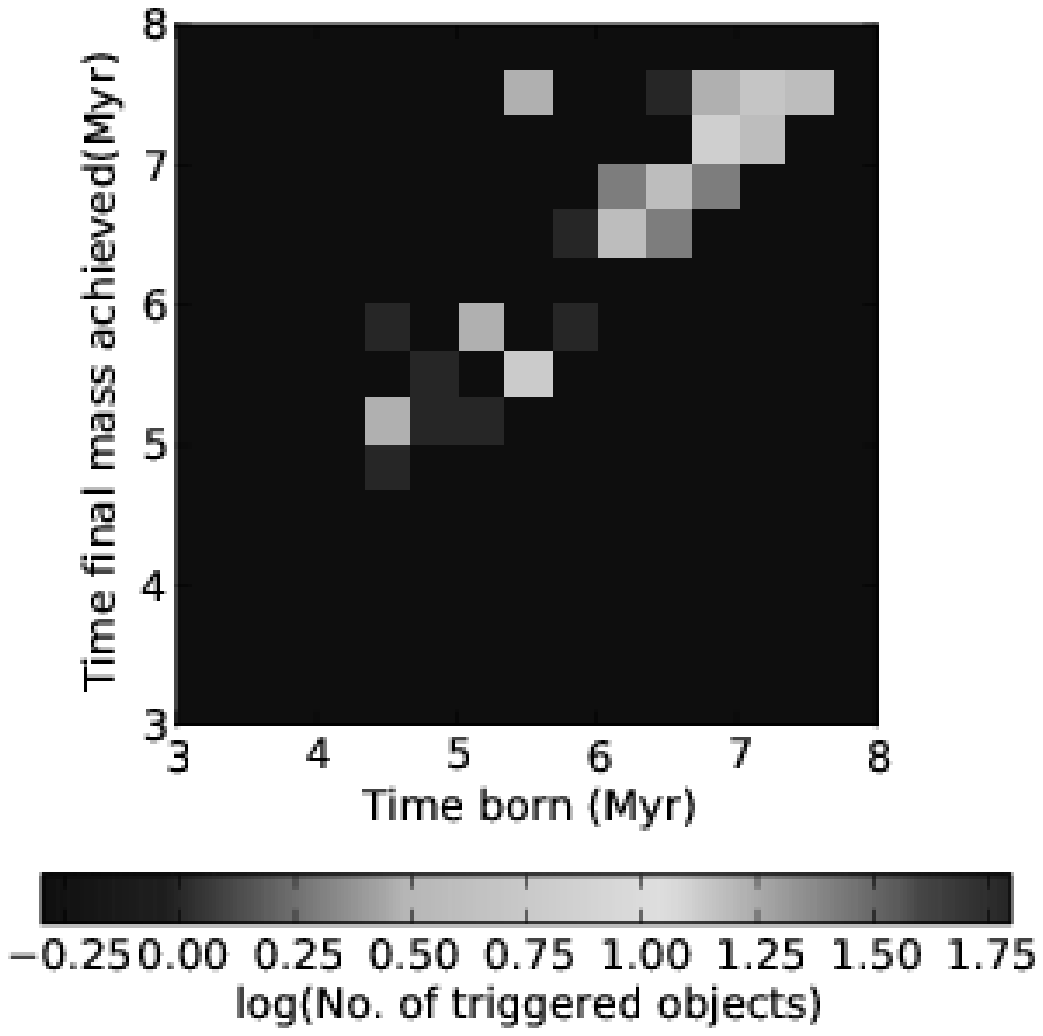}}
     \caption{Density plots illustrating in Run I the two--dimensional parameter space defined by the time at which a star was born (x-axis) and the time at which it achieved its final mass (y--axis) with colours representing the number of objects (panels a and c), the mean stellar mass (panels b and d) and the fraction of triggered objects (panel e) in each two--dimensional bin. The top row (panels a and b) refer to the control Run I and the bottom row (panels c, d and e) to the ionized Run I.}
   \label{fig:finals_runi}
\end{figure*}

\begin{figure*}
    \centering
     \subfloat[Run J control]{\includegraphics[width=0.34\textwidth]{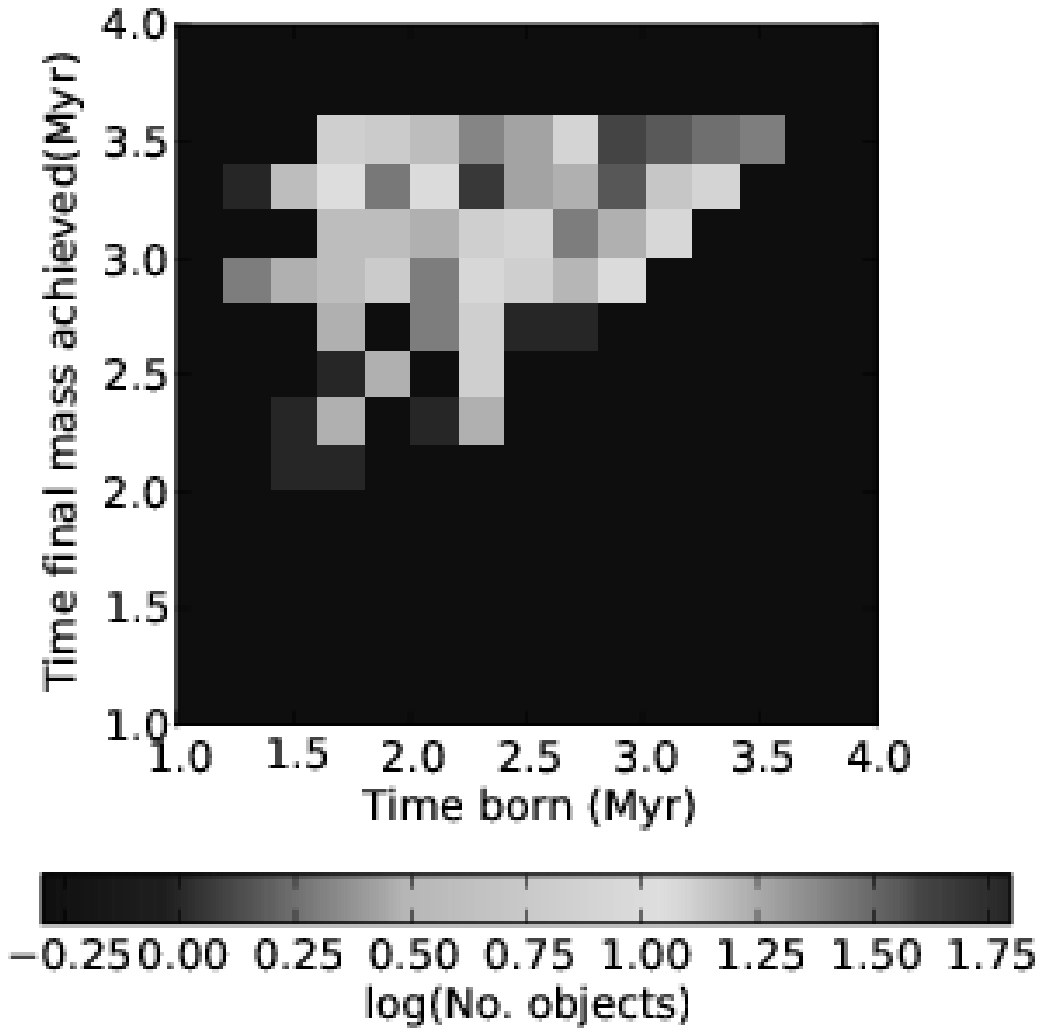}}     
     \hspace{.01in}
     \subfloat[Run J control]{\includegraphics[width=0.34\textwidth]{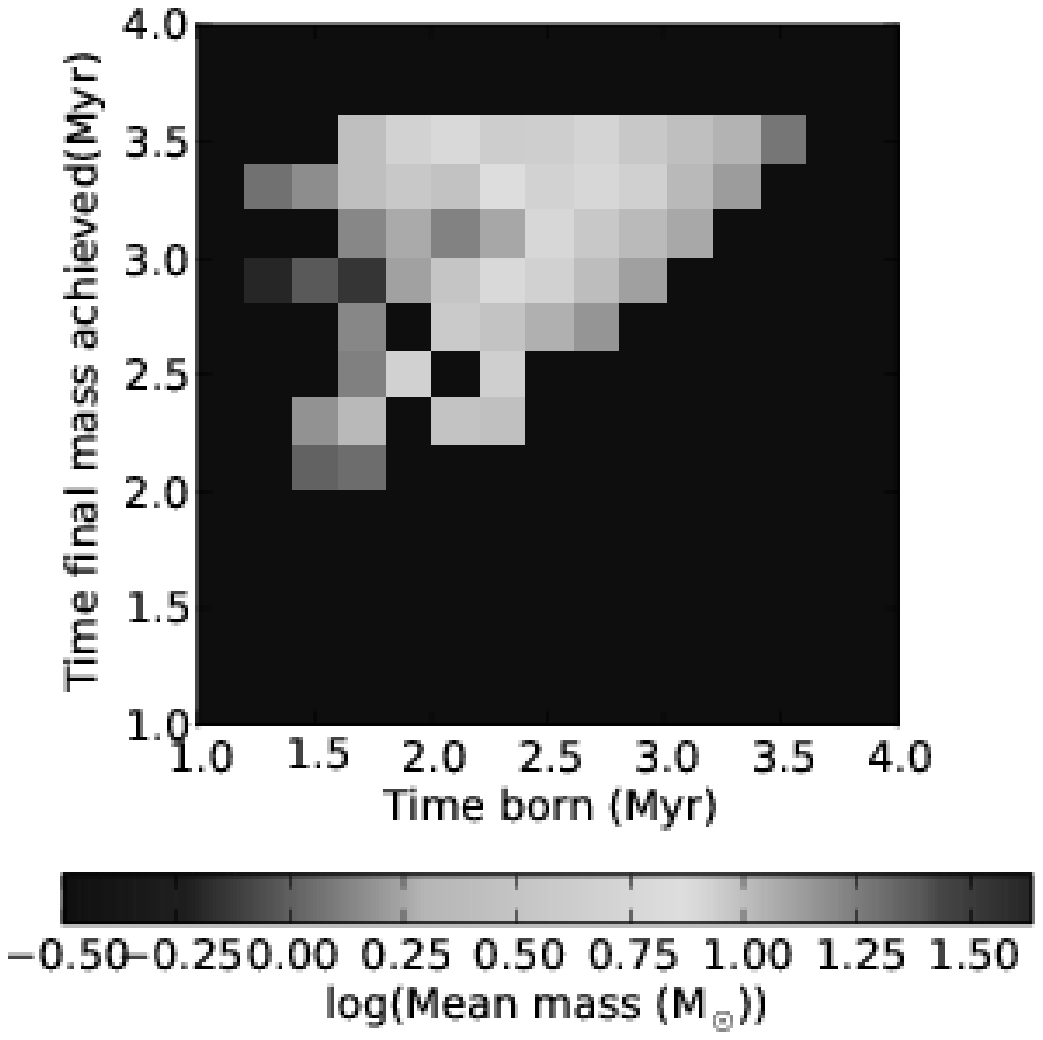}}
     \vspace{.01in}
     \subfloat[Run J ionized]{\includegraphics[width=0.32\textwidth]{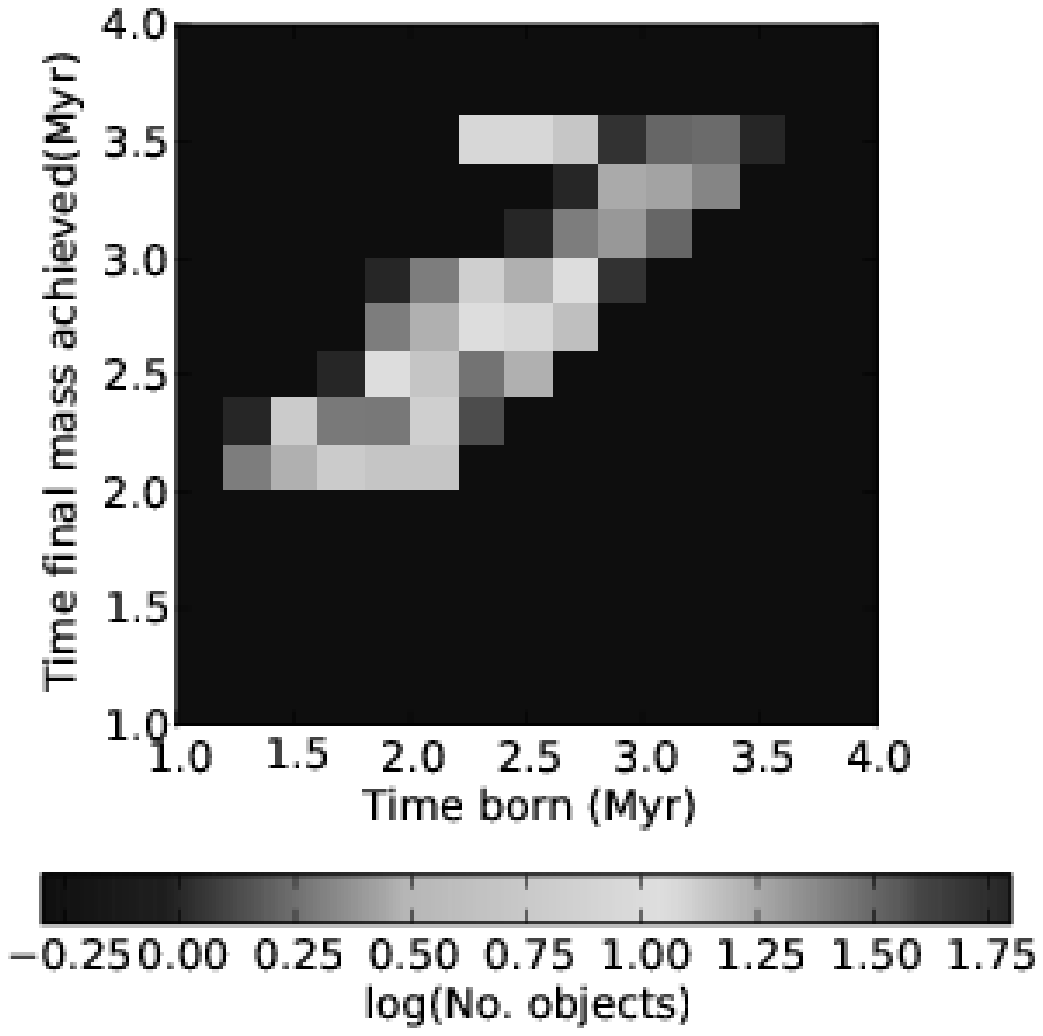}}     
     \hspace{.01in}
     \subfloat[Run J ionized]{\includegraphics[width=0.32\textwidth]{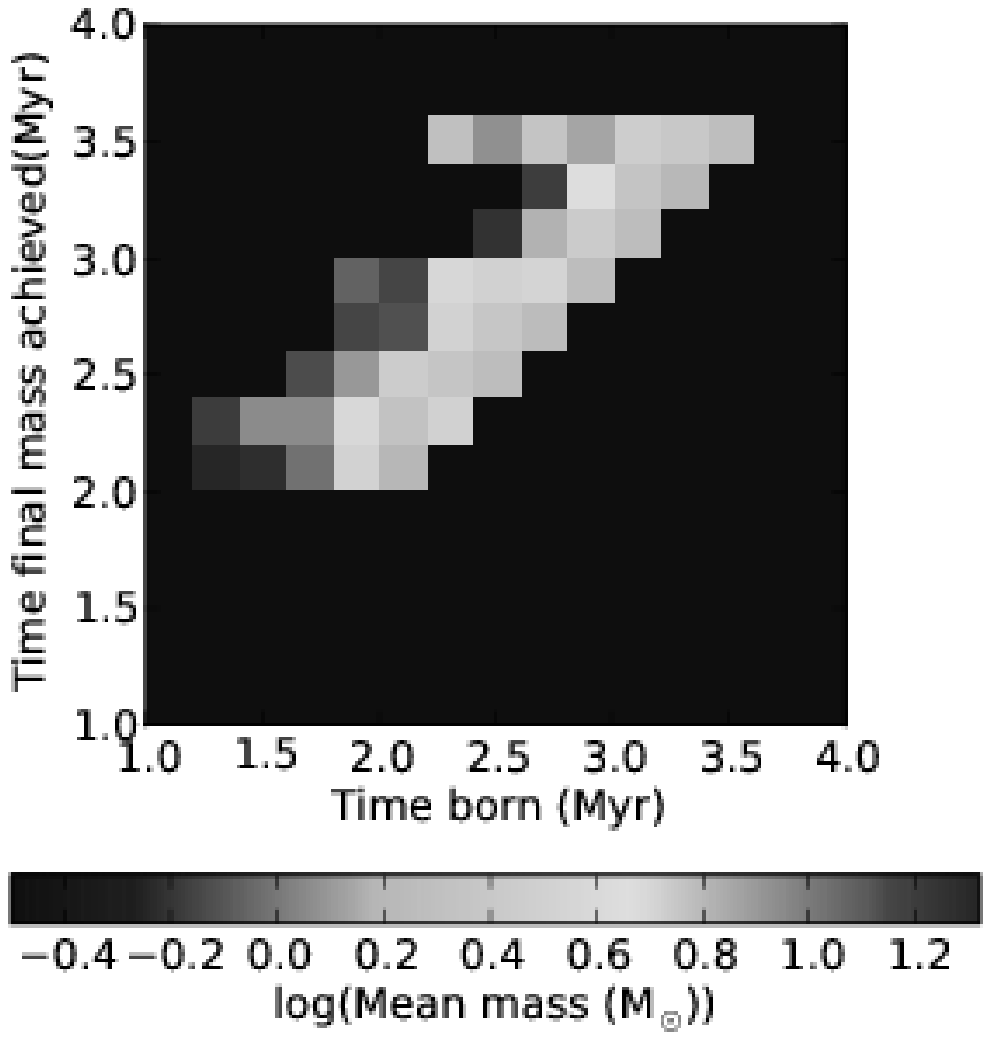}}
    \hspace{.01in}
     \subfloat[Run J ionized]{\includegraphics[width=0.32\textwidth]{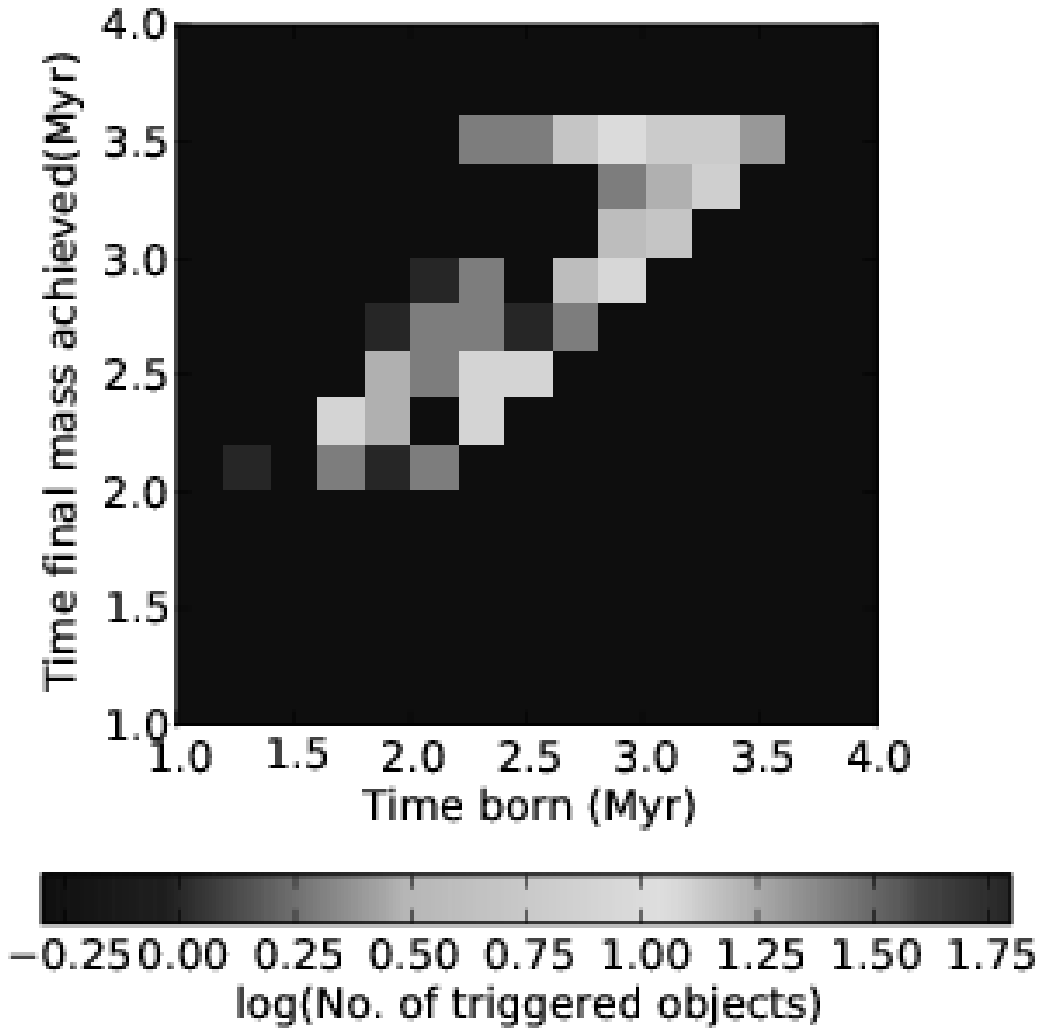}}
     \caption{Density plots illustrating in Run J the two--dimensional parameter space defined by the time at which a star was born (x-axis) and the time at which it achieved its final mass (y--axis) with colours representing the number of objects (panels a and c), the mean stellar mass (panels b and d), and the fraction of triggered objects (panel e) in each two--dimensional bin. The top row (panels a and b) refer to the control Run J and the bottom row (panels c, d and e) to the ionized Run J.}
   \label{fig:finals_runj}
\end{figure*}

\indent The concurrent changes in total stellar mass and total numbers of stars/clusters are reflected in changes in the stellar/cluster mass functions, depicted in Figure \ref{fig:compare_massfunc}. The mass functions of stars or clusters are plotted as overlaid semi--transparent histograms, where mass functions for the feedback runs are shown in green and those of the control runs are shown in blue. For Runs I and J, we also overlay a semitransparent red histogram showing the contribution of triggered objects to the mass functions in the ionized runs. In all cases, the feedback runs show depletions in higher--mass objects and excesses of lower--mass objects, although the statistics are clearly poor in Runs A and D. In the case of Run J, the shape of the mass function also appears different between the feedback run, which resembles a power--law with a turnover at low masses, and the control run, which is closer to log--normal in shape. The mass functions of the triggered objects appear to be of roughly the same shape as the all--star mass functions in the ionized runs, save that there are deficits of high--mass triggered objects. Since Runs I and J both contain $>$ 100 objects, it is legitimate to attempt a statistical analysis of their mass functions. In Figure \ref{fig:KS_runij}, we show cumulative distribution functions for the stars in these runs, blue being the mass function in the runs without feedback, green the mass function for all stars in the runs with feedback, and red the mass function of the triggered objects alone.\\
\indent We performed KS tests on the mass functions and found that the difference between the all--star control and ionized I runs is significant, with a 1.2$\%$ probability that these two mass functions are drawn from the same distribution, and that the difference between the two all--star J runs is highly significant, with a one part in $10^{-22}$ probability that the mass functions are from the same distribution. We also performed KS tests to determine if the mass functions of the triggered objects are significantly different from the all--star mass functions in the ionized runs, and found that they are not -- there are respectively 55$\%$ and 38$\%$ probabilities that the triggered and all--star populations in Runs I and J belong to the same populations. Since one of the effects of feedback is to slow or stop accretion onto the ionizing sources, we tested whether the apparent differences in the mass functions were due largely to differences at the high mass end. We repeated all of the above KS tests but excluded all stars with masses in excess of our minimum ionizing mass of 20M$_{\odot}$. We found that this did not affect the conclusions of the KS tests, implying that there are statistically--significant differences in the populations of low-- and intermediate--mass stars between the control and ionized runs.\\
\indent The reason for the stronger effect of triggering on the IMF in Run J can be deduced from Figures \ref{fig:finals_runi} and \ref{fig:finals_runj} which illustrate the stellar populations in the control and ionized Runs I and J distributed according to when the stars formed (on the x--axis) and when they achieved 95$\%$ of their final masses (on the y--axis) with colours representing the numbers of stars in each two--dimensional bin (panels a and c), the average mass of objects (panels b and d), and the numbers of triggered objects (panel e) in each two--dimensional bin. In both cases, the top row (panels a and b) refer to the control runs and the bottom row (panels c, d and e) refer to the ionized runs.\\
\indent As expected, in the control runs, most objects acquire their final masses towards the end of the simulations although they may form at any time, producing a strong horizontal feature towards the top of Figures \ref{fig:finals_runi}(a) and \ref{fig:finals_runj}(a). Most stars in the control runs, regardless of when they are born, continue accreting mass until the simulation stops. This is not true for all objects even in the control runs however, since some stars are dynamically ejected from dense gas and star--rich regions, or are born further out in the clouds and consume all the locally--available gas. In particular, in Run J, there are a few quite massive stars which are ejected early on from their parent subclusters by close encounters with even more massive objects, leading to few objects which form early, acquire their final masses early, but are relatively massive.\\
\indent In both ionized runs, the density plots in Figures \ref{fig:finals_runi} and \ref{fig:finals_runj} acquire a more diagonal shape, indicating that many stars achieve their final masses at roughly the same time they are born. In the ionized Run I, there is a clear population of stars, including the most massive, that reach their final masses early, at a simulation time of $\sim5$Myr, which is shortly after the ignition of the ionizing sources. These are stars in the central cluster which are starved of gas due to the destruction by the expanding HII regions of the accretion flows feeding the cluster. Excepting these objects, Figures \ref{fig:finals_runi}(a) and (c) are quite similar, indicating that most objects in the ionized Run I are able to continue accreting for most of the simulation. This is due to the fact that, as shown in Figure \ref{fig:trigIJ}, most of the stars in Run I not belonging to the central cluster (the majority of objects) are embedded in the accretion flows or the walls of the cavity excavated by the HII regions and still have access to dense gas. Figure \ref{fig:finals_runi}(e) shows that the a high proportion of objects which achieve their final masses soon after birth are triggered, indicating that the HII--region-driven shock triggers their formation, then washes over them and deprives them of further gas to accrete from.\\
\indent The differences between the control and ionized Run J simulations are stronger. Figures \ref{fig:finals_runj}(c) and (d) show that a large fraction of stars, forming at all times throughout the simulation, reach their final masses early after being deprived of material to accrete, and that these mostly end as low--mass stars. It is these objects which are responsible for the change in shape of the IMF in the ionized Run J relative to the control run. This behaviour is due to most objects in Run J occupying the central cluster, either because they formed there or have fallen there, which is largely cleared of gas by ionization, and Figure \ref{fig:finals_runj}(e) shows that there is no strong correlation between when stars form, when they acquire their final mass, and whether or not they are triggered, since even many of the triggered objects end up in the central cluster.\\
\indent In summary, Figures \ref{fig:finals_runi} and \ref{fig:finals_runj} clearly show that feedback truncates accretion and mass growth. Although there is significant triggering in Run I, most stars are able to accrete for most of the simulation and grow to intermediate or large masses and spread themselves out over the mass function. In Run J, there is also significant triggering but many stars are prevented from accreting to higher masses, which results in an excess of low--mass objects in the ionized run J compared to the control run.
\subsection{Structure of the clusters/associations}
\begin{figure*}
     \centering
     \subfloat[Run I]{\includegraphics[width=0.45\textwidth]{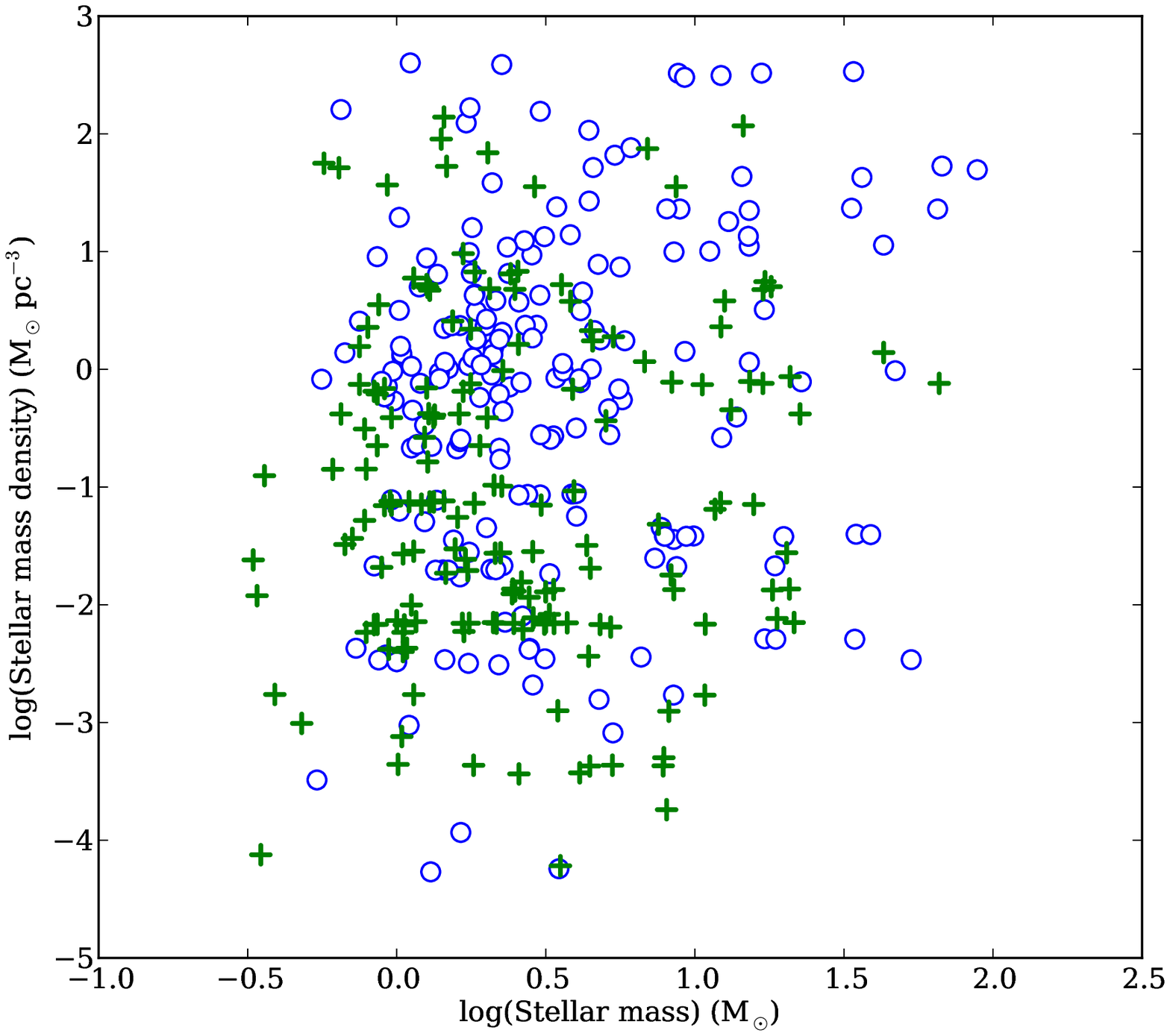}}     
     \hspace{.01in}
     \subfloat[Run J]{\includegraphics[width=0.45\textwidth]{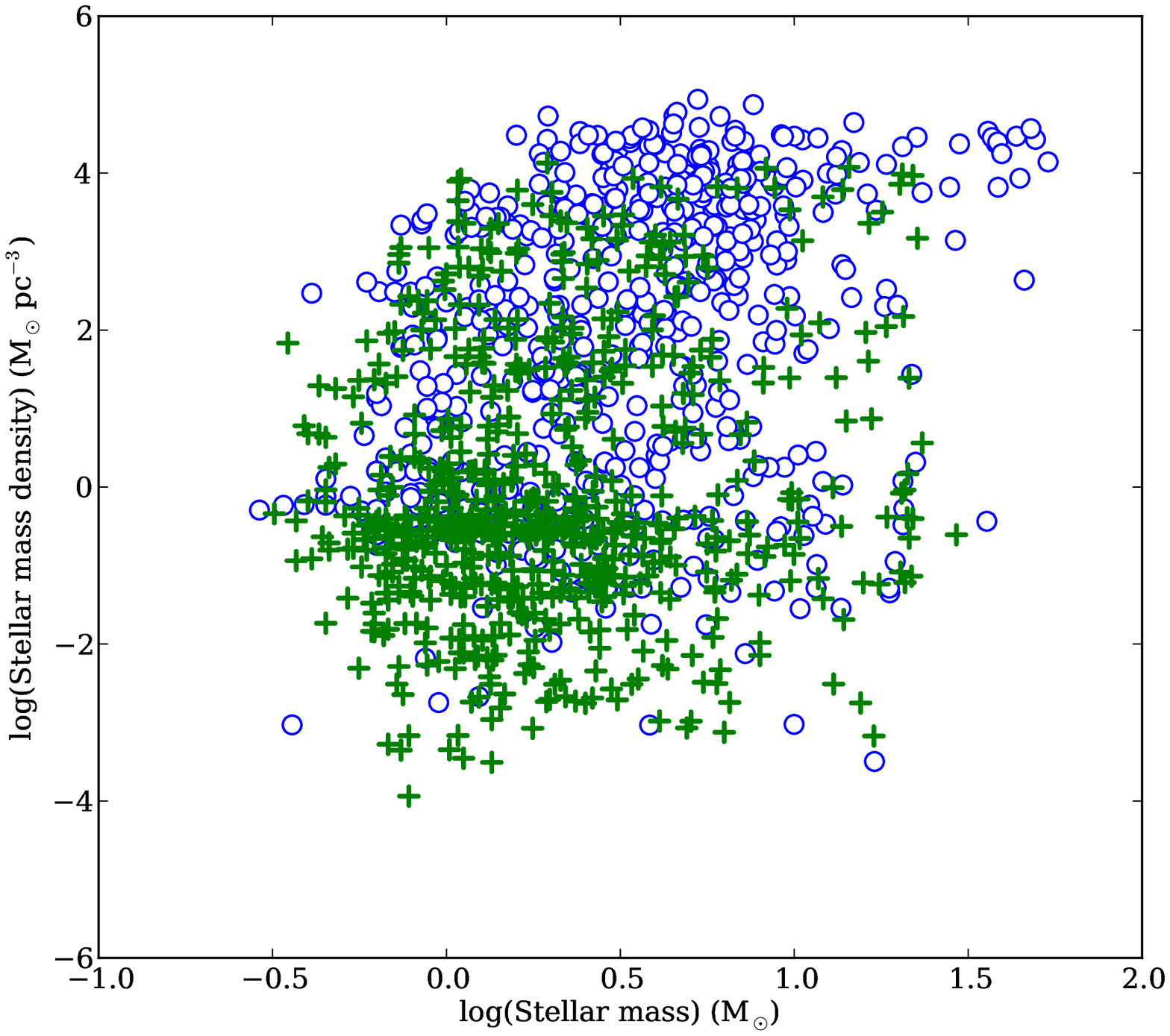}}
     \caption{Comparison of the local stellar mass density as a function of mass for every star in Runs I (left panel) and J (right panel) with stars from the control runs shown in blue circles and those from the ionized runs in green crosses.}
   \label{fig:sink_mass_dens}
\end{figure*}

\begin{figure*}     
     \subfloat[Run I]{\includegraphics[width=0.45\textwidth]{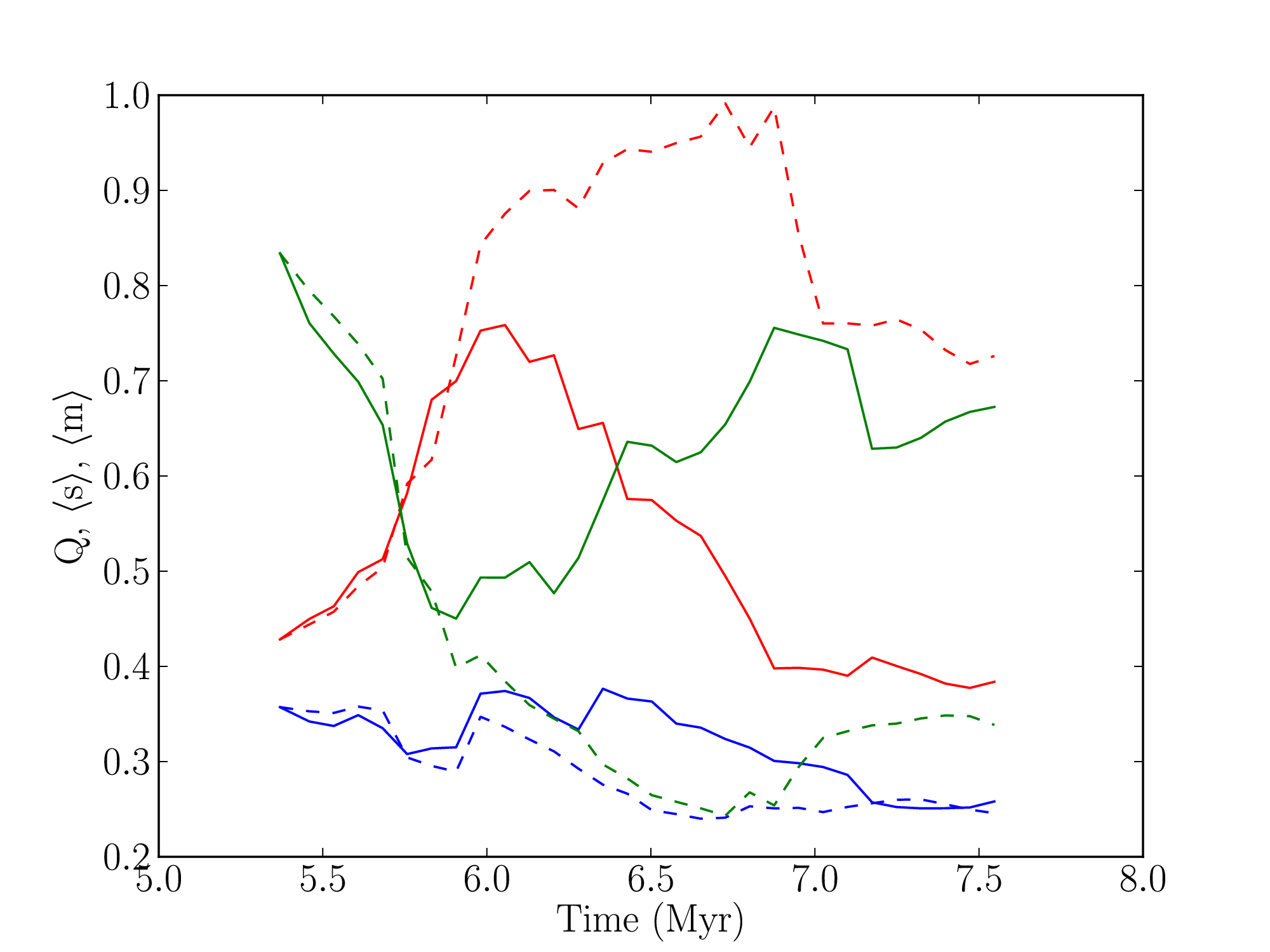}}     
     \hspace{.1in}
     \subfloat[Run J]{\includegraphics[width=0.45\textwidth]{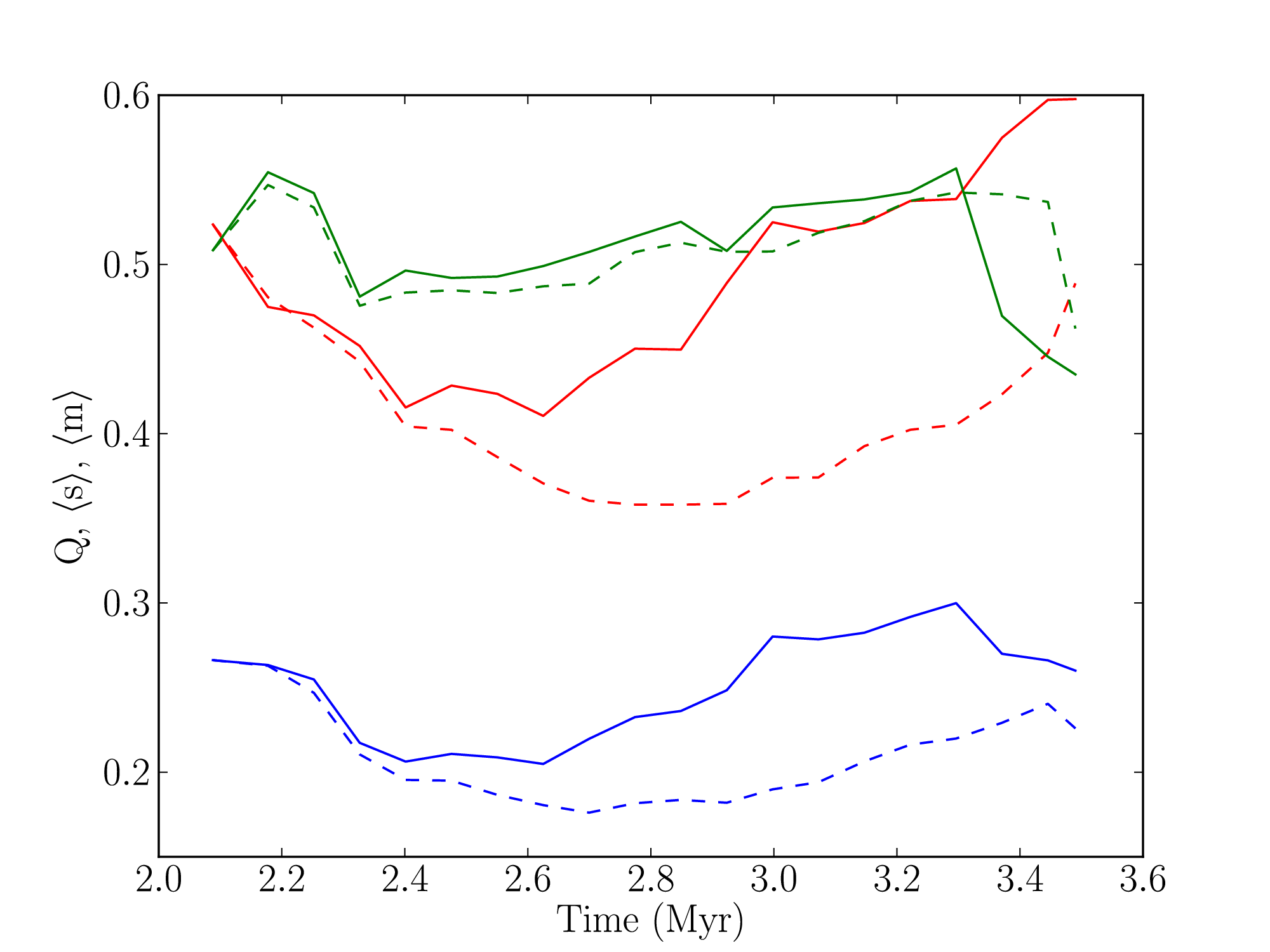}}
     \caption{Comparison of the time--evolution of the Q--parameter ($Q=\langle l \rangle/\langle s \rangle$, red lines), where $\langle s \rangle$ (green lines) denotes the mean separation between pairs of clusters and $\langle l \rangle$ denotes the mean edge--length of the minimum spanning tree connecting all clusters (blue lines), in the ionized (solid lines) and control (dashed lines) Runs I (left column) and D (right column). Q$<$0.8 indicates a fractally--substructured system, Q$>0.8$ indicates a smooth distribution of clusters with a global density gradient and Q=0.8 indicates a uniform distribution. We view both simulations down the down the x (top row), y (middle row) and z (bottom row) axes. In these simulations, the effect of feedback on the geometry of star--formation is strong, robust against changes of viewpoint, but affects the Q--parameter in opposite directions, decreasing it in Run I, resulting in a more fractal stellar distribution and doing the reverse in Run J.}
   \label{fig:compare_Q_IJ}
\end{figure*}

\indent It is clear from Figure \ref{fig:compare_end} and Figure \ref{fig:trigIJ} that feedback, as well as changing the numbers and masses of stars relative to control runs, can also produce clusters/associations with very different geometries. In particular, in Runs I and J, the major groupings of stars appear to be less concentrated in the ionized runs, due to gas expulsion, and star formation is more widely distributed throughout the cloud volumes due to triggering. This can be quantified by measuring the stellar mass density in the neighbourhood of each star in the two runs. We estimate this by finding the radius around each star which contains exactly ten other stars and compute the local stellar mass density from the total mass within the sphere so defined, ignoring gas. The results are depicted in Figure \ref{fig:sink_mass_dens}. In all cases there is a great deal of scatter but it is clear that the feedback--influenced runs exhibit lower stellar mass densities, particularly for the massive stars, and that this effect is stronger in Run J. Feedback therefore tends to limit the formation of very rich clusters and the resulting systems are substantially more vulnerable to external tidal disruption.\\
\indent It is also clear that the geometrical distribution of stars in the control runs is also quite different, however. In order to make quantitative statements on this subject, an objective means of characterizing the structure of a group of stars or clusters is required. We make use of the Q--parameter devised by \cite{2004MNRAS.348..589C}. The parameter is defined for a set of points, usually in projection as
\begin{eqnarray}
Q=\frac{\langle l \rangle}{\langle s \rangle},
\end{eqnarray}
where $\langle s \rangle$ denotes the average separation between two points, and $\langle l \rangle$ denotes the mean edge--length of the minimum spanning tree that uniquely connects the points. The Q--parameter is useful because it can distinguish between distributions which are fractal or subclustered (Q$<$0.8, smaller values indicating more subclustering) and distributions which are smooth but have global density gradients (Q$>$0.8, larger values indicating steeper density gradients). A Q--value of 0.8 would correspond to a uniform distribution of stars.\\
\indent We find that none of these statistics are strongly altered by feedback in Runs A or D and do not discuss these runs further here. In Figure \ref{fig:compare_Q_IJ}, we plot the statistics for Runs I and J. Since the minimum--spanning--tree edges and stellar separations are projected quantities, we checked to see whether these was any significant change in these plots if a different viewing axis were chosen, but we found that there was not. There are significant differences between the control I run and the control J run, and between each control run and the corresponding feedback run.\\
\indent In the Run I control simulation, most of the stars are to be found in the central cluster which forms at the junction of the filaments in the gas. As the cluster continues to accrete and grow more populous, dynamical interactions smooth out any substructure and it evolves towards a Q--value of between 0.8 and 1.0, indicating a smooth distribution with a modest density gradient. Towards the end of the simulation, some stars are ejected, which has the effect of making the cluster appear slightly more substructured, decreasing its Q--value to just below 0.8. Conversely, in the control Run J, the higher average densities and stronger shocks lead to more distributed and subclustered star--formation, mostly associated with the dense filaments in the gas. As star--formation proceeds, the Q--value thus declines to being with, but mergers of the subclusters towards the latter half of the simulation begin to erase the substructure and the Q--value beings to climb again.\\
\indent In the ionized Run I, triggered star--formation at the tips of the accretion flows produces several tightly clustered groups of stars outside the main cluster inhabited by most objects in the control run. This leads to a small increase in $\langle l \rangle$ (shown in blue), a much larger increase in $\langle s \rangle$ (shown in green), and a corresponding decrease in Q (shown in red), indicating a \emph{more} substructured system. Instead in the ionized Run J, feedback generates numerous, rather uniformly--distributed triggered stars outside the main cluster, as well as smoothing out the main cluster itself. This results in a statistically smoother cluster because the shell of triggered stars and the smearing out of the central cluster partially increases $\langle s \rangle$ by a larger factor than it increases $\langle l \rangle$, thus increasing Q relative to the control run. This results in the feedback--influenced system being \emph{less} substructured and more like a single cluster with a density gradient.\\
\section{Morphological indications of triggering}
\begin{figure}
\includegraphics[width=0.5\textwidth]{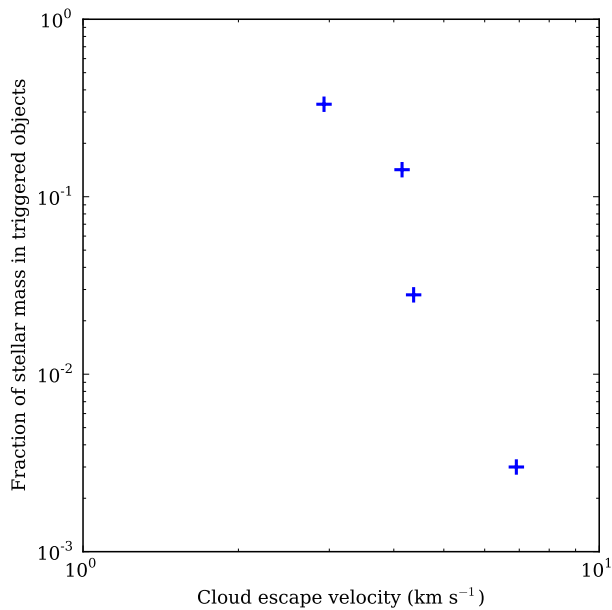}
\caption{Mass of triggered objects as a fraction of total stellar mass as a function of escape velocity for the four clouds considered here.} 
\label{fig:trig_vesc}
\end{figure}

\indent The overall effect of feedback on the star--formation process in all the simulations presented here is negative, as measured by the star--formation efficiency, although there is evidence for at least some triggered star or cluster formation in all the calculations. This is in contrast to the findings of \cite{2007MNRAS.375.1291D} and \cite{2012MNRAS.tmp.2723D}, where we saw respectively a modest increase, and no change in star--formation efficiency due to external feedback. External feedback acts, at least initially, on the outskirts of clouds in regions where there is little or no star formation. If it influences the star--formation process at all, it is likely to be by driving some of this quiescent gas into the denser regions of clouds, which may increase the star--formation efficiency. By contrast, internal feedback acts in regions where star formation is at its most vigorous and expels potentially star--forming gas from these regions. We find, however, that the \emph{numbers} of stars/clusters formed are increased by feedback in three of the four pairs of simulations and that the stellar mass functions are altered by feedback in a statistically--significant fashion, becoming biased towards lower masses. This is again in contrast to the findings of \cite{2012MNRAS.tmp.2723D} and the difference can again be attributed to the much more direct action of internal feedback on the regions of clouds which are already forming stars in these calculations.\\
\indent Two of the most commonly--cited visual signposts of triggered star formation are expanding shells where young stars are embedded in dense gas on the borders of feedback--driven bubbles, and pillars of dense cold gas hosting young stars and pointing towards the massive stars which sculpted them. We see these features in abundance in our simulations. The gas structure in Run I is initially rather simple, with a few accretion flows funnelling gas into a central cluster. In the control run, this structure simply persists over the $\sim$3Myr since this is comparable to the freefall time in the system and the gas is therefore far from being exhausted. In contrast, the gas structure in the feedback I run is rather more complicated, consisting (in z--projection) of two large bubble structures and a pillar several pc long pointing towards the central ionizing cluster. It is very clear from examining Figure \ref{fig:compare_end} that the bubbles have been blown in directions where there were no accretion flows, and that the pillar visible in the z--axis projection is itself the remains of one of these flows. Pillar--like structures are often thought to be caused by dense clumps of material shielding a roughly conical region of a cloud from a point radiation source but these results suggest that at least some pillars may instead be the remains of coherent flows of material feeding gas into cluster potential wells. In hindsight, it is rather natural that this should be so. While the tip of the accretion flow is destroyed by ionization and a photoevaporation--driven shock is created just beyond, the rest of the flow continues to funnel gas towards where the tip of the flow once was, so gas collects behind the shock where the accretion flow is being ionized. This hypothesis should be relatively easy to test observationally, since it would imply that most of the material in a pillar, and particularly the pillar interior, would be moving collectively \emph{towards} the source of ionizing radiation, at least early on in the evolution of the HII region.\\
\indent We used our ability to trace the different fates of parcels of gas with or without the influence of feedback to unequivocally identify triggered and non--triggered stars or clusters. Applying this technique to Runs I and J reveals reasonably strong, but by no means perfect, correlation between triggered stars and shell-- or pillar--like structures in the gas. In Figure \ref{fig:trigIJ} (e), we see that the objects nearest the tip of the very prominent pillar visible in Run I are in fact \emph{not} triggered (although objects further down the column of the pillar are). This is easy to understand based on our strict definition of triggering which identifies objects as triggered only if the gas from which they form would not otherwise be involved in star formation. The pillar in Run I is the remains of an accretion flow and the material from which the three stars near its tip form is, in the control run, delivered by this flow to the central cluster where it is involved in star formation. It is also clear from Figure \ref{fig:trigIJ} that the association of stars with bubble walls is a good but not foolproof indicator of triggering, since bubbles may overrun regions which were already in the process of forming stars or sweep such regions up and incorporate them into bubble walls. Despite these counterexamples, it is clear that the gas morphology and spatial distribution of stars in the ionized Runs I and J is dominated by the action of feedback so, at least in these environments, these features do provide good indicators that triggered star formation is underway, but they do not necessarily aid in the identification of which stars are triggered and which are not.\\
\indent In Paper I, we postulated that the main determinant of the the efficacy of feedback across our parameter--space of clouds was the cloud escape velocity, since the sound speed in the HII regions is fixed. We plot in Figure \ref{fig:trig_vesc} the mass involved in the formation of triggered objects as a fraction of the total mass involved in star formation in the four clouds considered here and find a relation which reasonably follows a power law, which seems to support this hypothesis.\\
\indent Feedback also alters the geometrical distribution of stars in some of our model clusters, but in rather different ways. In the ionised Run I, the triggering of what are essentially subclusters at the heads of the accretion flows which were feeding gas into the cloud's central cluster produces a more fractal distribution of stars and a very low Q value. This is in contrast to the control Run I where the unimpeded formation of a central cluster with few outliers leads to an almost uniform morphology (i.e. a Q--value close to 0.8). In Run J on the other hand, both the control and ionized runs have significant fractal substructure due to the higher initial gas densities and more distributed and vigorous star formation in these calculations. In this case, feedback triggers star formation in some otherwise quiescent regions at large distances from the central concentration of clusters, and smears out this central concentration itself, leading to a geometrically smoother distribution of stars and a higher value of Q. Both of these conclusions are robust against projection effects.\\
\indent From these findings, it seems difficult to infer any general indicators of triggered star formation purely from the geometrical distribution of stars in a given system, whether the masses of the stars are taken into account or not. We find that the triggered formation of whole clusters (in Runs A and D) is rather rare and that the geometrical distribution of the clusters which do form is not strongly affected by ionization feedback. Taking into account also the velocities of stars or clusters (three--dimensional or projected) does not aid in the identification of triggered objects either, since such objects are well--mixed with spontaneously--formed stars or clusters in velocity--space. It is possible that other feedback mechanisms, particularly supernovae, may have a stronger effect at the extra--cluster scale.
\section{Conclusions}
\indent We have investigated the effects of ionizing feedback from massive stars on the star--formation process in four model molecular clouds known from Paper I to be significantly influenced by feedback. We find that, although there is strong evidence of triggered star formation and even of occasional triggered cluster formation, the overall effect on the star--formation efficiency is always to decrease it. The expulsion of potentially--star forming gas and the disruption of accretion flows feeding objects that have already formed always outweighs triggering in these calculations.\\
\indent Feedback of this form is able to significantly influence the stellar IMF, either by simply shifting it to lower masses by curtailing accretion onto all stars, as mostly seen in Run I,  or by forming an excess of new stars and preventing them from accreting, as seen in Run J. Feedback also limits the growth of the most massive stars. However, the IMFs so formed are scarcely unusual in appearance and there is no obvious way of inferring the presence or degree of triggering purely from observing the mass function in a given system.\\
\indent Similarly, although we find that ionizing feedback can profoundly alter the geometrical distribution of the stars within a cluster in ways that are robust against projection effects, the degree to which, and the direction in which, it does so depend on the cluster environment. It is not the case that internally--triggered star formation always leads to more subclustering (as measured quantitatively by the Q parameter), since it may produce a relatively smooth halo of triggered stars around an otherwise structured central cluster. In Run I, the density is low so that it takes a long time for the expanding bubbles to sweep up shells which are gravitationally unstable -- most of the triggering in Run I on the few Myr timescales considered here therefore happens in the accretion flows which dominate the cloud's structure. In Run J, the background density is higher and the sweeping up of material by expanding shells efficiently produces widely distributed triggered stars. Feedback may thus increase or decrease the Q parameter depending on the properties of the background cloud.\\
\indent Despite the difficulty of inferring triggering from the geometry of the stars alone, the \emph{combined} geometry of the stars and the surviving cold gas provides an indication of where induced star formation is occurring, by the association of triggered stars with shells or pillars, but even this correlation is not always reliable. We suggest that the partial destruction of accretion flows feeding dense gas towards ionizing sources may be a natural explanation for pillar--like structures.\\
\indent Triggering of star formation by ionization is seen in our simulations, but does not compensate for the decrease in star formation due to gas expulsion by the same process. Discerning which stars are triggered remains problematic and their overall numbers are only marginally significant compared to those generated by ongoing spontaneous star formation. In particular, we find that it is not not always safe to infer that individual stars have been triggered merely from their association with pillars or bubble walls.
\section{Acknowledgements}
We thank the anonymous referee for a very careful reading of the paper and for several suggestions that helped improve its readability.

\bibliography{myrefs}

\label{lastpage}

\end{document}